\documentclass[preprint,a4paper,12pt,onecolumn,twopage]{elsarticle}

\usepackage{physics}
\usepackage[ampersand]{easylist}
\usepackage{tabularx,tabulary}
\usepackage[hidelinks,hypertexnames=false]{hyperref}
\usepackage[export]{adjustbox}
\hypersetup{
	colorlinks,
	citecolor=blue,
	filecolor=black,
	linkcolor=blue,
	urlcolor=black}
\usepackage{url}
\usepackage{csquotes}

\usepackage{algorithm}
\usepackage{algpseudocode}
\usepackage[toc,page]{appendix}
\usepackage{graphicx}
\usepackage{amsthm,amsfonts,amsmath,amssymb}
\usepackage{mathrsfs,bm,bbm}
\usepackage{enumerate}
\usepackage{enumitem}
\usepackage{graphicx}
\usepackage{subfig}
\usepackage[noabbrev]{cleveref}
\usepackage{float}
\usepackage{color}

\usepackage{standalone}
\usepackage{varwidth}
\usepackage{xfrac}
\usepackage{makecell}
\usepackage{cancel}
\usepackage{pgfplots}
\usepackage{pgfplotstable}
\pgfplotsset{compat=newest}

\usetikzlibrary{calc,shapes,shapes.geometric,arrows, arrows.meta,decorations.pathmorphing,patterns,backgrounds,pgfplots.colorbrewer}
\usetikzlibrary{shapes,arrows.meta,calc,fit,backgrounds,shapes.multipart,positioning}
\newdimen\XCoord
\newdimen\YCoord

\pgfdeclarelayer{background}
\pgfdeclarelayer{foreground}
\pgfsetlayers{background,main,foreground}

\pgfplotsset{every axis/.style={scale only axis}}
\pgfplotsset{
	legend entry/.initial=,
	every axis plot post/.code={%
		\pgfkeysgetvalue{/pgfplots/legend entry}\tempValue
		\ifx\tempValue\empty
		\pgfkeysalso{/pgfplots/forget plot}%
		\else
		\expandafter\addlegendentry\expandafter{\tempValue}%
		\fi
	},
}

\pgfplotsset{
	cycle list/.define={my marks}{
		every mark/.append style={solid,fill=\pgfkeysvalueof{/pgfplots/mark list fill}},mark=*\\
		every mark/.append style={solid,fill=\pgfkeysvalueof{/pgfplots/mark list fill}},mark=square*\\
		every mark/.append style={solid,fill=\pgfkeysvalueof{/pgfplots/mark list fill}},mark=triangle*\\
		every mark/.append style={solid,fill=\pgfkeysvalueof{/pgfplots/mark list fill}},mark=diamond*\\
	},
}

\definecolor{c1}{RGB}{27,158,119}
\definecolor{c2}{RGB}{217,95,2}
\definecolor{c3}{RGB}{117,112,179}
\definecolor{c4}{RGB}{231,41,138}

\tikzstyle{startstop} = [rectangle, rounded corners, minimum width=3cm, minimum height=1cm,text centered, draw=black, fill=black!50]
\tikzstyle{i} = [trapezium, trapezium left angle=70, trapezium right angle=110, minimum width=3cm, minimum height=1cm, text centered, text width=7cm, draw=black, fill=red!30]

\tikzstyle{o} = [trapezium, trapezium left angle=70, trapezium right angle=110, minimum width=3cm, minimum height=1cm, text centered, text width=7cm, draw=black, fill=green!30]

\tikzstyle{process} = [rectangle, minimum width=3cm, minimum height=1cm,text centered, text width=4.5cm ,  draw=black, fill=black!10]
\tikzstyle{decision} = [diamond, minimum width=3cm, minimum height=0.5cm, text centered, inner sep=1pt, draw=black, fill=black!30]

\tikzstyle{joint} = [draw=black,circle,node distance=3cm,fill=yellow!20]

\tikzstyle{arrow} = [ultra thick,->,>=stealth]

\tikzset{
	connector/.style = {draw,circle,minimum width=1cm, minimum height=1cm, text centered,fill=yellow!20},
}
\tikzset{block/.style={rectangle split, draw, rectangle split parts=2, text badly centered, text width = 4cm, font=\fontsize{10}{0}\selectfont},   
	line/.style={draw, -{Latex[length=2mm,width=1mm]}},
	cloud/.style={draw, ellipse,fill=white!20, node distance=3cm, minimum height=4em},  
	container/.style={draw, rectangle,dashed,inner sep=0.28cm, rounded
		corners,fill=yellow!20,minimum height=4cm}}

\newsavebox{\measurebox}
\definecolor{editc}{RGB}{0,0,255}
\definecolor{colora}{rgb}{0.93,0.77,0.77}
\definecolor{colorb}{rgb}{0.85,0.93,0.77}
\definecolor{colorc}{rgb}{0.77,0.93,0.93}
\definecolor{colord}{rgb}{0.85,0.77,0.93}

\newtheorem{remark}{Remark}[section]
\newtheorem{theorem}{Theorem}
\DeclareRobustCommand{\T}{\intercal}
\crefname{equation}{}{}
\allowdisplaybreaks

\usepackage{subfig}
\biboptions{sort&compress}
\usepackage{nomencl}
\makenomenclature

\journal{Applied Energy}

\begin{document}

\begin{frontmatter}


\title{A Secure Distributed Ledger for Transactive Energy: The Electron Volt Exchange (EVE) Blockchain}



\author{Shammya Saha \corref{cor1}\fnref{fn1}}
\author{Nikhil Ravi \fnref{fn1}}
\author{K\'{a}ri Hreinsson \fnref{fn1}}
\author{Jaejong Baek \fnref{fn3}}
\author{Anna Scaglione \fnref{fn1}}
\author{Nathan G. Johnson \fnref{fn5}}
\address{Arizona State University, Arizona, USA}

\cortext[cor1]{Corresponding author, email: shammya.saha@asu.edu}
\fntext[fn1]{Shammya Saha, Nikhil Ravi, K\'{a}ri Hreinsson, and Anna Scaglione are with the School of Electrical, Computer and Energy Engineering, Arizona State University, Tempe, AZ, USA.}
\fntext[fn3]{Jaejong Baek is with Laboratory of Security Engineering For Future Computing and with Center for Cyber-Security and Digital Forensics, Arizona State University, Tempe, AZ, USA.
}
\fntext[fn5]{Nathan G. Johnson is with The Polytechnic School, Arizona State University, Mesa, AZ, USA.}

\begin{abstract}
The adoption of blockchain for Transactive Energy has gained significant momentum as it allows mutually non-trusting agents to trade energy services in a trustless energy market. Research to date has assumed that the built-in Byzantine Fault Tolerance in recording transactions in a ledger is sufficient to ensure integrity. Such work must be extended to address security gaps including random bilateral transactions that do not guarantee reliable and efficient market operation, and market participants having incentives to cheat when reporting actual production/consumption figures. Work herein introduces the Electron Volt Exchange framework with the following characteristics: 1) a distributed protocol for pricing and scheduling prosumers' production/consumption while keeping constraints and bids private, and 2) a distributed algorithm to prevent theft that verifies prosumers' compliance to scheduled transactions using information from grid sensors (such as smart meters) and mitigates the impact of false data injection attacks. Flexibility and robustness of the approach are demonstrated through simulation and implementation using Hyperledger Fabric.
\end{abstract}


\end{frontmatter}

\mbox{}
\nomenclature[01]{$\mathcal{A}$}{Calligraphic letters are sets}
\nomenclature[02]{$\abs{\mathcal{A}}$}{Denotes the cardinality of set $\mathcal{A}$}
\nomenclature[03]{$T$}{Number of intervals in the decision horizon}
\nomenclature[04]{$\mathcal{N}$}{Set of aggregators}
\nomenclature[05]{$N$}{Number of aggregators}
\nomenclature[06]{$\mathcal{B}$}{Set of buses/nodes $b$ in the power grid}
\nomenclature[07]{$\mathcal{B}^{(n)}$}{Subset of buses managed by aggregator $n\in\mathcal{N}$}
\nomenclature[08]{$\mathcal{G}_e $}{Electric grid graph}
\nomenclature[09]{$\mathcal{G}_c $}{Communication network graph}
\nomenclature[10]{$\mathcal{E}_e$}{Set of edges/lines connecting the set of buses $\mathcal{B}$ in $\mathcal{G}_e$}
\nomenclature[11]{$\mathcal{E}_c$}{Set of communication links between aggregators $\mathcal{N}$}
\nomenclature[12]{$\mathcal{T}$}{Set of time intervals in the decision horizon}
\nomenclature[13]{$GL$}{Global ledger}
\nomenclature[14]{$LL_n$}{Local ledger for aggregator $n \in \mathcal{N}$}
\nomenclature[15]{$\bm x$}{Boldfaced lower-case letters denote vectors and $x_i$ denotes the $i^{\textnormal{th}}$ element of a vector $\bm{x}$.}
\nomenclature[16]{$\bm X$}{Boldfaced upper-case letters denote matrices and  $X_{ij}$ denotes the $ij^{\textnormal{th}}$ entry of a matrix $\bm{X}$}
\nomenclature[17]{$\bm X^{\T}$}{Transpose is denoted by $^\T$, so $\bm X^{\T}$ is the transpose of $\bm X$}
\nomenclature[18]{$\bm{X}^\dagger$}{Is the pseudo-inverse of $\bm{X}$, where $\bm{X}^\dagger=(\bm{X}^\T\bm{X})^{-1}\bm{X}^\T$}
\printnomenclature


\section{Introduction}
\label{intro}
The proliferation of Distributed Energy Resources (DERs), Electric Vehicles (EVs), grid-level energy storage, and networked grid-edge devices requires a trustworthy open energy trading platform for participants -- i.e., a Transactive Energy (TE) framework. TE combines financial signals and dynamic control techniques to shift the timing and quantity of energy usage to achieve greater efficiency, increased use of renewable energy, reduced energy costs, and improved flexibility to manage shifts in net load locally. Such benefits have motivated the increasing body of research whose goal is to manage real-time demand and electricity supply in an open market where prosumers and utilities interact to establish a market-clearing price. Examples are the auction mechanisms proposed in \cite{LIN2019113687}, algorithms for co-simulation of transmission and distribution networks combined markets \cite{NGUYEN2019666}, multi-agent models capturing trading behaviors  \cite{JANKO2018715}, and thermostatically controlled loads to participate in  TE markets \cite{ BEHBOODI20181310}, to name a few examples. 

Recently, many researchers have purported blockchain as the ideal enabling platform to implement TE. Blockchain can enhance cyber-security and traceability of Peer-to-Peer (P2P) transactions between mutually non-trusting parties in the TE marketplace~\cite{VANLEEUWEN2020114613}. There are several benefits to such an implementation: 1) once stored on the ledger, all transactions are transparent to all participants through an identical copy of the ledger, 2) new transactions are \textit{hash-chained} when appended to the ledger, an operation that makes them immutable, mitigating cyber-attacks aimed at reducing the integrity and availability of the data, and 3) all functional aspects of TE enabled by blockchain, from bidding to pricing to billing, can be orchestrated running \emph{Smart Contracts} \cite{Mohanta2018}. 

\subsection{Contribution}
\begin{figure}
	\centering
	\includegraphics[width=0.8\linewidth]{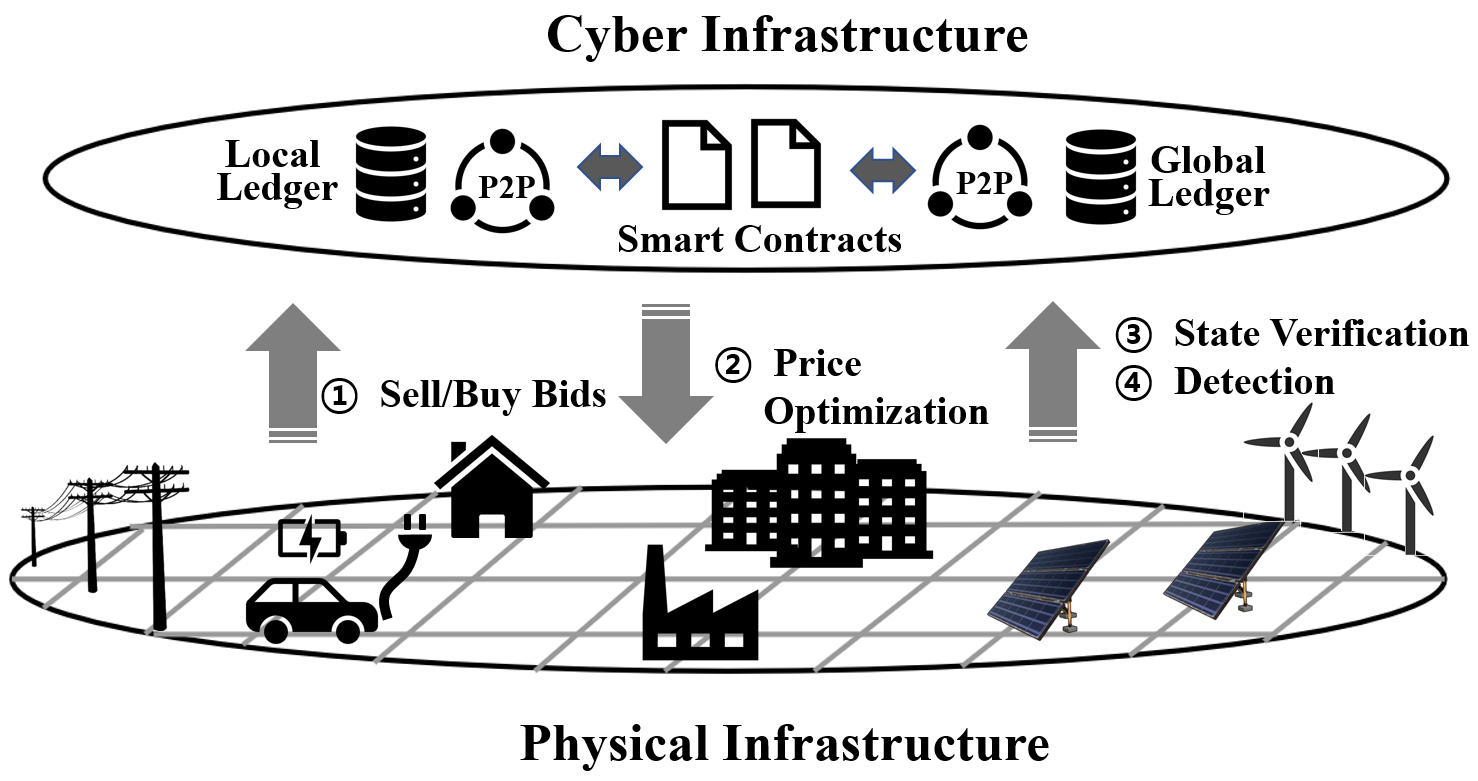}
	\caption{Conceptual architecture of EVE illustrating cyber and physical layers.}
	\label{fig:concept}
\end{figure}

Blockchain ensures transparency and immutability of bidding and trading records in the ledger. However, records in the ledger have security gaps during the submission of bids and the verification of contractual obligations. In fact:
\begin{enumerate}[leftmargin=*]
    \item Threats exist internal to TE approaches because selfish players have an intrinsic incentive to cheat on reported consumption needs or production capacity during market clearing \cite{Hussain2019}.
    \item Ex-post, if the market stakeholders control smart meters, they can inject false data to hide discrepancies that would otherwise reveal cheating. 
\end{enumerate}
To address these issues we propose the \textbf{Electron Volt Exchange (EVE)} blockchain architecture (\Cref{fig:concept}). Novelty compared to other TE blockchain research lies in the following components:
\begin{enumerate}[leftmargin=*]
    \item TE blockchain designs commonly consider only bilateral transactions. Instead, the EVE approach utilizes a decentralized solution for the entire market economic dispatch problem whose formulation falls in the class of network utility maximization problems, first proposed for real time pricing in \cite{li2011optimal} (\Cref{sec:pricingAlgo}). The closest to our approach is found in \cite{munsing2017blockchains}, where the authors have incorporated controllable loads and generation to develop an iterative pricing algorithm using a smart contract that updates global variables of the distributed optimal power flow problem. The scheme still relies on a central update of variables to achieve convergence. Compared to \cite{munsing2017blockchains}, this work incorporates renewable generation, thermostatically controlled loads (TCLs), storage devices, deferrable appliances (DAs), and electric vehicles (EVs) and relies on a hierarchical, distributed architecture including aggregators \cite{Rosell2018} to \textquote{divide and conquer} the communication problem, avoiding congestion and yielding a scalable implementation for optimal price calculation.
    \item The EVE architecture includes the first blockchain-based, distributed Robust State Verification (RSV) mechanism for TE transactions, where physical sensor measurements are cross-validated in a decentralized fashion to ensure prosumers abide by their market commitment (\Cref{sec:verification}). Our algorithm, inspired by the work \cite{vukovic2014security} on distributed state estimation in adversarial settings, is shown to be robust against False Data Injection Attacks (FDIAs) aimed at TE market theft.
    \item The pricing and verification algorithms are tested via numerical simulations in \Cref{sec:simulations}. Implementation of the EVE blockchain framework onto a distributed ledger \cite{hlf_paper} is described in \Cref{sec:imp} using the open-source Hyperledger Fabric (HLF) framework. It includes a customized BFT-SMART \cite{sousa2018byzantine} consensus protocol to provide security and improve performance with Byzantine Fault Tolerance (BFT) in an untrustworthy environment. This improves upon standard security features of HLF, such as Membership Service Provider (MSP), Fabric CA (Certificate Authority), Attribute-Based Access Control (ABAC) \cite{Yuan_abac}, and others that address common security concerns. \textcolor{black}{Bench-marking for the proposed smart contracts using Hyperledger Caliper \cite{hlf_wp} is also described herein.} 
\end{enumerate}

\subsection{Related Work}
\label{sec:literature}
There is a substantial body of research and industrial efforts in TE; we focus our literature review on the relatively recent trend that includes blockchain as the backbone for managing P2P communications for market-related operations. These operations involve handling and securing prosumer bids and dispatch values, deciding a market-clearing algorithm, ensuring the balance between demand and supply to meet network constraints, ensuring cyber-security, and more. \textcolor{black}{Readers are referred to \cite{Musleh2019} for a broad discussion on the applications of blockchain for smart grid cyber-physical infrastructure, to \cite{Mylrea2017} for potential benefits of blockchain for grid resiliency against cyber-attacks, and to \cite{WU2021106593} for details on how smart inverter, advanced metering infrastructure, and energy coordinator can support the digitization and decentralization of TE.} 

Existing literature has focused on 
\textcolor{black}{co-simulation of the physical grid in tandem with blockchain TE implementation \cite{HAYES2020105419}}, quantifying energy losses caused by energy transactions in an energy blockchain \cite{Sanseverino2017}, enforcing proportional fairness among DERs participating in voltage control through smart contracts to ensure voltage stability \cite{danzi2017distributed}, \textcolor{black}{developing a blockchain based energy trading platform for electric vehicles \cite{Silva2019}}, integrating energy and carbon markets through a blockchain based trading framework \cite{HUA2020}, and more. The two-layer blockchain implementation in \cite{Lu2019} consists of a first layer with smart meters forming a private blockchain and a second layer with aggregators forming a consortium blockchain to coordinate energy transactions within regions. These works assume security is implicit in transactions between aggregators and flexible demand assets. The authors in \cite{Zhao2018} use a continuous double auction mechanism, but the transaction mechanism exposed the prosumer's unique ID, posing a privacy concern. \textcolor{black}{Similar concerns described in \cite{AGUNG2020} indicate a public blockchain exposes the transactions and balances of each prosumer, and further, that the rate of transaction processing limits scalability.} Authors in~\cite{Kang2017} present a double auction mechanism for localized energy exchanges between EVs, where local aggregators publicly audit and share transaction records without relying on a trusted third party. The use of local aggregators also appears in \cite{Ferreira2018} to create an energy market that combines blockchain and IoT for two flexible community market players: an EV community and a DER community. In \cite{munsing2017blockchains}, the authors propose a blockchain implementation to clear the market using the Alternative Direction Method of Multipliers (ADMM) in a master-slave distributed architecture, where a central aggregator/master node updates global variables. The multi-layer smart contract implementation based on Ethereum in~\cite{Danzi2018} addresses the mismatch in the settlement between \textit{System Operators} and \textit{Balance Responsible Parties}, yet it does not decentralize the Independent System Operator (ISO) nor include verification. Work done in \cite{Jogunola2019} uses Hyperledger Composer to demonstrate a blockchain based TE market implementation while excluding a market clearing price calculation and assuming energy transactions are automatically verified. A prototype implementation of a TE market using Hyperledger Fabric for metering and billing purposes is proposed in \cite{Gur2019}, yet pricing and verification were not addressed. A related study that uses Hyperledger Composer in~\cite{Saifur2018} determines the market clearing price by averaging bid prices offered by all buyers while sorting sellers by first-in, first-out basis. \textcolor{black}{An approach similar to \cite{Saifur2018} using an Ethereum based blockchain architecture is found in \cite{CHRISTIDIS2021115963}.} However, such algorithms can be easily exploited by malicious prosumers or attackers to manipulate the clearing price and destabilize the TE market. Malicious prosumers can similarly influence the co-simulation framework presented in \cite{Jonathan2018} where the ratio of total generation and consumption reported by the prosumers is used to determine the price. \textcolor{black}{The private blockchain solution proposed in \cite{Zheng2019} requires a match between the energy producer and consumer regarding the amount of power to be generated and consumed, respectively, which is not practical for many prosumers and can violate physical constraints on the distribution network. The blockchain based energy trading model in \cite{ZHANG2020} also lacks sufficient protection from physical constraint violations because the approach allows for an open trading platform across diverse types of power sellers without any optimized market pricing.} In most studies, Smart Contracts orchestrate information exchanges among participants and during recording transactions, while still requiring a central entity to be in charge of calculating the market-clearing price in contrast to our fully decentralized solution. Also, prior methods focusing on security aspects and countermeasures against cyber-attacks to the market-clearing mechanism have ignored physical verification that must be tied to market records to work effectively as a continuous deterrent to theft. Relevant to this study are also prior works integrating smart metering with blockchain, specifically energy trading applications~\cite{pop2018blockchain}. Even if they leverage the immutability of blockchain, these approaches leave data integrity and privacy concerns unresolved \cite{ANDONI2019143}.  

To address these gaps, our work incorporates insights from the considerable body of research that has been developed in cyber-security of electric power measurement systems to encompass stealth FDIAs in state estimation~\cite{xie2010false}, non-stealth state estimation attacks such as data jamming~\cite{deka2015optimal}, bias injection attack \cite{LUO2019113703}, and denial of service attacks~\cite{vukovic2014security}. To the best of our knowledge, the only work that discusses possible attack scenarios in blockchain-based energy trading is~\cite{Wang2019}; the scenarios mentioned by the authors include a malicious stakeholder attempting to modify market operations to produce an inaccurate clearing price, a malicious market operator attempting to modify operations of the market algorithm, and a malicious outsider trying to remotely tamper with communications among TE market participants. The RSV presented in this work addresses these security concerns to enhance the blockchain-based TE framework's cyber-security. The market modeling approach in~\cite{Wang2019} uses blockchain only to collect bid information from the prosumers and then utilizes a centralized architecture for determining the market clearing price. On the contrary, this work uses blockchain to manage prosumers, while the proposed decentralized price optimization algorithm uses the decentralized architecture of blockchain to run the iterative price determination algorithm. \textcolor{black}{Moreover, using the inherent security features of Hyperledger Fabric and ABAC allows EVE to avoid the complex attribute based encryption for transaction security introduced in \cite{GUAN202134} while still achieving the same level of privacy.}
\section{EVE as a Cyber-Physical System Architecture}
\label{sec:cps}\label{sec:TE}
\Cref{fig:concept} represents the interactions tied to the TE application layer for a generic blockchain based cyber-physical infrastructure implementation. The specifics of our architecture are summarized in the following sections.
\subsection{Physical Infrastructure}
The physical infrastructure includes: 
\begin{itemize}[leftmargin=*]
  \item The electrical grid modeled as a connected graph $\mathcal{G}_e=(\mathcal{B},\mathcal{E}_e)$. Lines and transformers are characterized by admittance parameters $y_{ij},~\forall~ij\in \mathcal{E}_e$\footnote{Voltage control and protective equipment are ignored because these do not have a direct impact on market operation and hence are not required for the description of the EVE architecture.}.
  \item Market Participants (prosumers and aggregators).
  \item Electrical loads, distributed generation, and storage assets that connect to buses on the electrical network. For clarity in our formulations, we model a single prosumer per bus $b$, making them equivalent\footnote{In cases where multiple consumers connect to the same physical bus, we model those as separate buses connected through zero impedance edges, but omit this in diagrams for simplicity.}.
  \item Electrical sensors and control equipment.
\end{itemize}

\begin{figure}[ht]
  \centering
  \subfloat[]{\includegraphics[width=0.5\linewidth,height=0.38\linewidth]{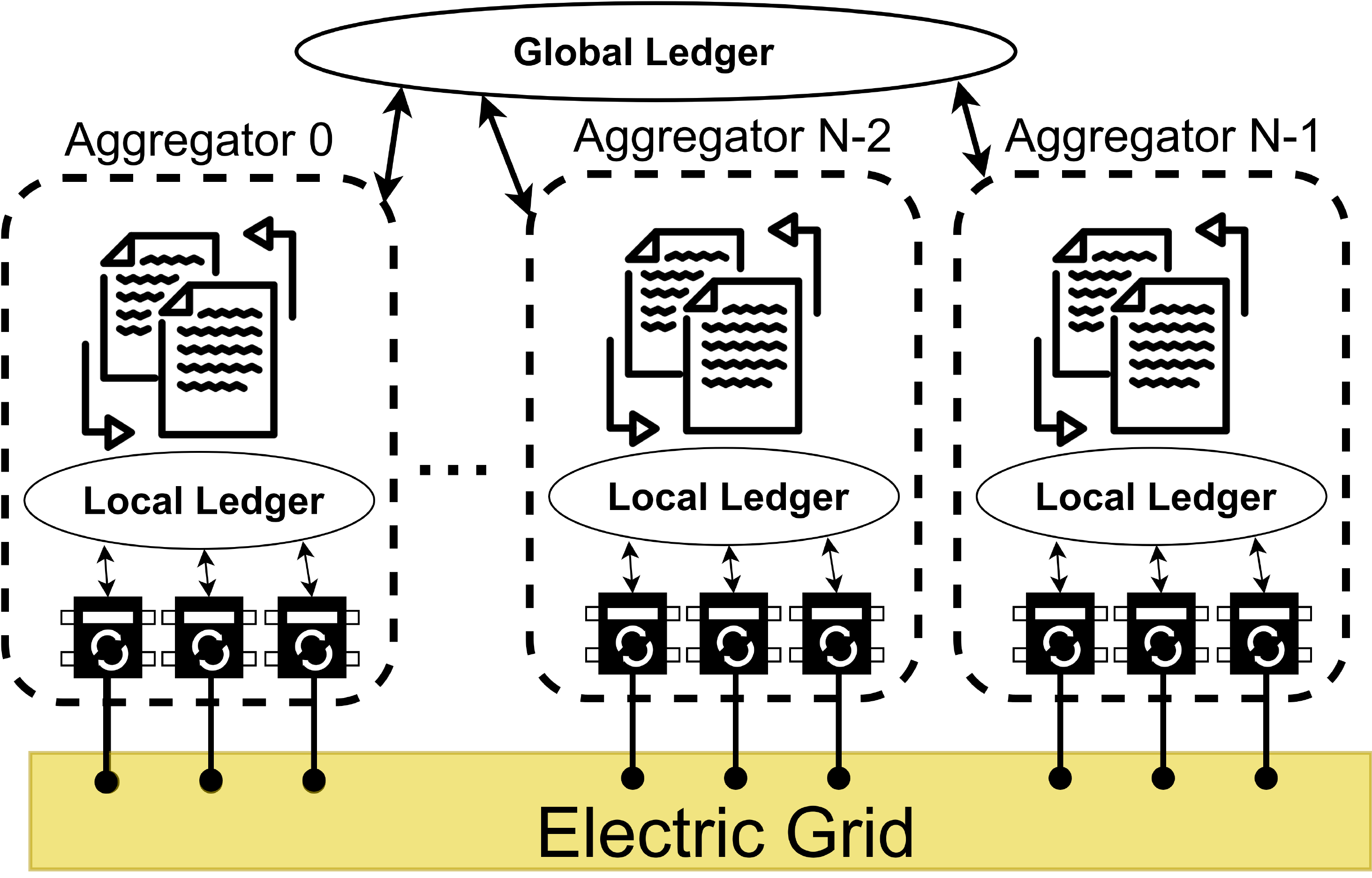}\label{fig:GG-LL}}
  \hfill
  \subfloat[]{\includegraphics[width=0.40\linewidth,rotate=90,keepaspectratio]{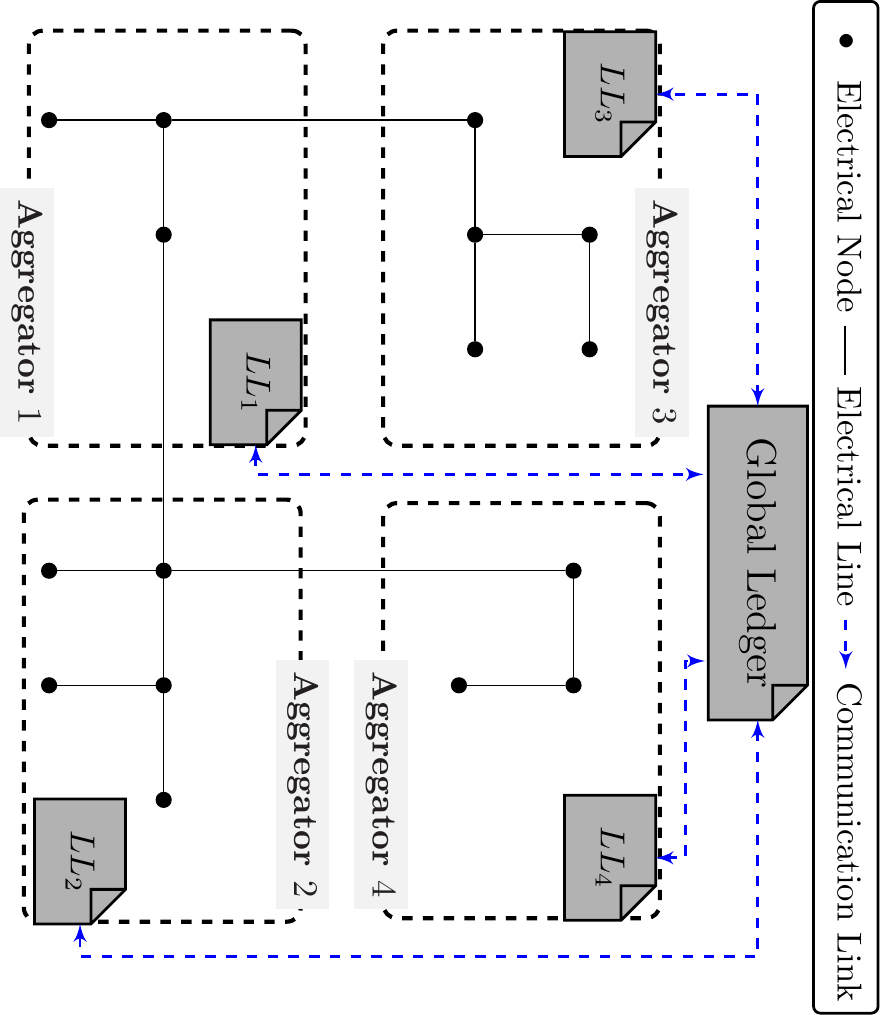}\label{fig:hiearchy}}
  \caption{\protect\subref{fig:GG-LL} The hierarchical distributed ledger architecture.  \protect\subref{fig:hiearchy} Cyber infrastructure overlaid with the physical structure for a sample distribution network with 4 aggregators.}
\end{figure}

\subsection{Application Layer}
\label{sec:application_layer}
In EVE, scalability is achieved by dividing the application layer entities into agents with distinct tasks.
\begin{itemize}[leftmargin=*]
    \item The bottom layer includes \textit{prosumers} who can buy and/or sell energy and control flexible loads, storage, and generation assets connected to the physical network. This layer can be broken into several physical regions. 
    \item The top layer includes a set of local \emph{aggregators} ($\mathcal{N}$) managing the prosumers connected to a subset of buses $\mathcal{B}^{(n)}\subset \mathcal{B}$ for all $~n\in \mathcal{N}$. 
\end{itemize}
The following remark is in order:
\begin{remark}
Individual prosumers have traditionally been unable to participate in energy markets, but aggregators may have access through the lumped capacity bids \cite{Mahmoudi2014}\footnote{A practical example can be found in \cite{calcca}.}. \textcolor{black}{Aggregators can act as intermediaries between small consumers/producers and volatile markets, and thereby provide hedging solutions to reduce risk to individual market participants \cite{KOCH201535}. Aggregators can procure demands from consumers and sell to purchasers through trading frameworks proposed in prior works \cite{6938882}.} Moreover, resources needed for price optimization and verification processes can be more easily be obtained and justified for aggregators rather than each prosumer. Aggregators do not complicate management because blockchain is suited for a decentralized architecture. A solution without aggregators would otherwise increase communication latency and computation time for market-clearing as new prosumers are added, reducing scalability.
\end{remark}

Policies in EVE can be divided into three main classes:
\begin{itemize}[leftmargin=*]
    \item \textbf{Pricing}: Policies deciding the optimum prosumers schedule and the price.
    \item \textbf{Verification}: Policies processing sensor information to verify that load/generation is correctly reported, contractual obligations have been met and billed accordingly.
    \item \textbf{Billing}: Policies for billing and ensuring compliance.
\end{itemize}

\subsection{Cyber Infrastructure}
The cyber infrastructure includes security policies for settling transactions, communication/computation resources, and data archival based on blockchain. Building the cyber network requires all market participants to work together as a consortium using a set of policies agreed to during network initialization to determine the participants' permissions. For the shared database or ledger within the cyber architecture, EVE uses CouchDB~\cite{couchdb} as it supports rich queries when data values are modeled as JSON. The cyber framework is generic to include or exclude Transport Layer Security (TLS); however, we recommend including TLS for additional security. 

Application of the policies mentioned in \Cref{sec:application_layer} is handled through distributed ledgers. In EVE, a ledger consists of (a) a database that holds current values of a set of ledger states, and (b) a transaction log that records all changes that have resulted in the current system state. Our implementation of EVE consists of two types of distributed ledgers, a Local Ledger ($LL$) and a Global Ledger ($GL$). The smart contracts in this work handle interactions between the ledgers ($GL$ and $LL$) and external applications to complete every transaction within EVE. \Cref{fig:GG-LL} shows the hierarchical architecture of those ledgers whereas \Cref{fig:hiearchy} overlays the distribution of physical nodes into aggregator zones. An aggregator uses the $LL$ to collect bids submitted by prosumers, verify local state information, and update individual prosumer budgets after verification. Aggregators access the $GL$ for distributed pricing and verification algorithms and for sharing global information with other aggregators. All information exchanges and history between participants are handled through smart contracts for reading, writing, and storing in distributed ledgers. 

\begin{figure*}[htbp]
	\centering
	\includegraphics[width=\textwidth]{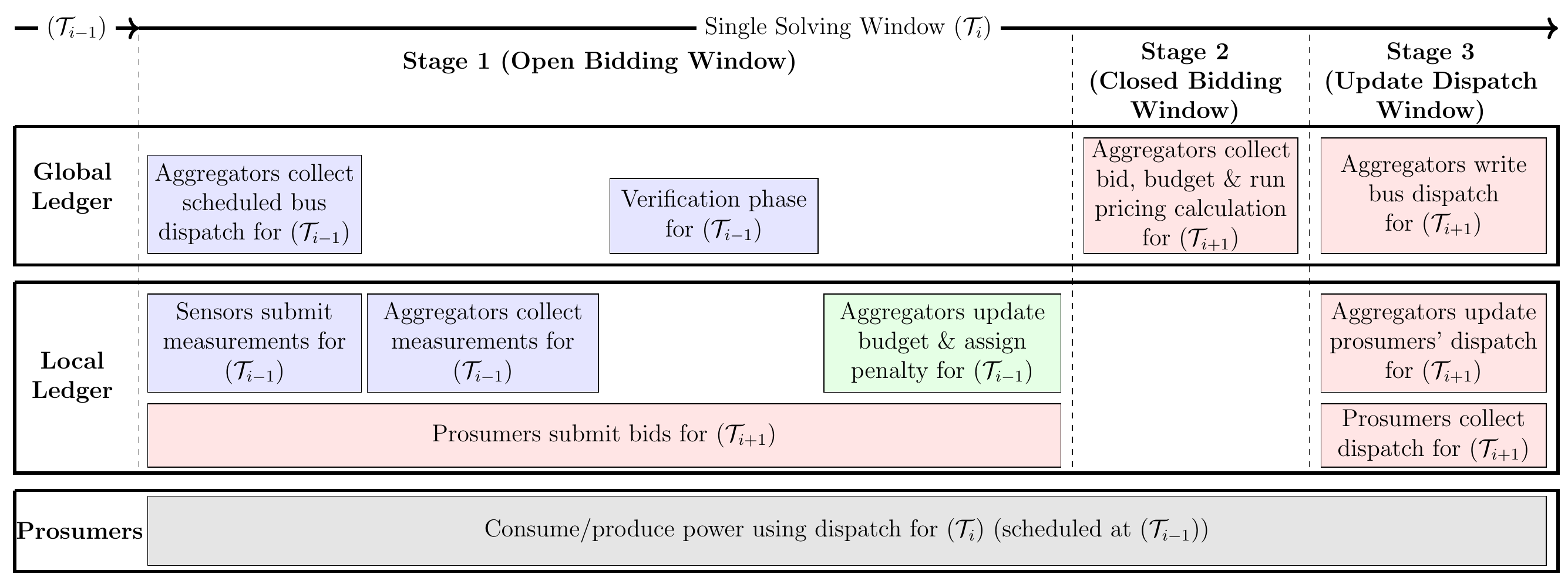}
	\caption{Overview of EVE major tasks and timeline for a single solution window with red, blue, and green used to illustrate Pricing, Verification, and Billing policy execution steps, respectively.}
	\label{fig:timeline}
\end{figure*}

\Cref{fig:timeline} depicts the implementation of policies under a single solving window $\mathcal{T}_i$ divided into three stages. Stage 1 refers to the Open Bidding Window in which prosumers submit bids for $\mathcal{T}_{i+1}$. Aggregators execute the verification algorithm based on measurements collected for $\mathcal{T}_{i-1}$. Hence stage 1 includes execution of the verification policy and then the billing policy. Stage 2 refers to the Closed Bidding Window in which the aggregators execute the distributed pricing algorithm for $\mathcal{T}_{i+1}$. Stage 3 refers to the Update Dispatch Window in which aggregators update the prosumers' schedules. 

In \Cref{sec:pricingAlgo} and \Cref{sec:verification} the paper details the pricing algorithm and verification policies respectively, along with mechanisms that ensure the integrity of the decision-making process. The numerical performance via simulations in \Cref{sec:simulations} and the implementation of the cyber architecture on HLF described in \Cref{sec:imp} conclude the paper.

\section{EVE Distributed Pricing Algorithm}
\label{sec:pricingAlgo}
Pricing and scheduling decisions are illustrated for a single solving window, as shown in \Cref{fig:timeline}. The period is split into $T$ smaller discrete intervals of unit duration, all in the set $\mathcal{T}_i=\{iT,\ldots,(i\!+\! 1)T\!-\!1\}$. Within each period, $\mathcal{T}_i$, bus $b$ is connected to assets that either supply or demand power. Net real power generation of bus $b$ at time $t$ is denoted by $p_b(t)$. Positive and negative values of $p_b(t)$ indicate that bus $b$ is supplying power to the grid or consuming from the grid, respectively. Neglecting losses, the total power schedule managed by aggregator $n\in\mathcal{N}$, $p^{(n)}(t)$, is the sum of power from each individual bus $p_b(t)$ associated with it:
\begin{align}
  \label{eq.sum-inj}
  p^{(n)}(t) &=\textstyle{\sum_{b\in \mathcal{B}^{(n)}}}  p_b(t).  
\end{align}

Each component $p_b(t)$ must have a certain cost (disutility) and must satisfy a set of constraints that depend on the generation and storage capacity available at the supply side, as explained in \Cref{sec.demand}, that determines the optimal schedule. In \Cref{sec.distributed}, we describe a decentralized dual decomposition algorithm to solve a power balancing problem between different aggregators with a dual variable reflecting the price of energy. Related works (\cite{li2011optimal,chang2012coordinated}) use a similar dual decomposition algorithm to solve a distributed problem between an aggregator and its customers, with the former reflecting its internal energy procurement cost function through iterative retail pricing and the latter trying to minimize deviation from a pre-determined aggregate power profile. Our work differs in that we are only modeling a decentralized energy market mechanism (hence no central provider) while leveraging the distributed ledger to ensure the liquidity of purchasers.

\subsection{Flexible Resource Model}\label{sec.demand}
The power injection trajectory $\bm{p}_b=[p_b(iT),\ldots,p_b((i+1)T-1)]^\T\in {\mathbb R}^{T\times 1}$ is constrained depending on the type of energy service bus $b$ provides. For example, in response to price signals, a participant may accept shifting to a less comfortable thermostat reference temperature, defer use of the dishwasher, dim lights, or other actions. This section describes in general terms\footnote{Specific examples are given in \ref{app:drmodels}.} the inter-temporal constraints for demand and supply in a single period $\mathcal{T}_i$ ($i$ is omitted for brevity) and discusses associated cost functions $C_b(\bm{p}_b)$. 

For each aggregator $n$ and bus $b$, the load profile can be split into a flexible component, that changes based on price, and an inflexible one that prosumers are willing to buy at any price. In the literature, $\bm{p}_b$ is typically modeled by a linear, affine function of a corresponding control signal $\bm{u}_b$, i.e.,\ for all $b \in \mathcal{B}^{(n)}, n \in \mathcal{N}$:
\begin{equation}
  \bm{p}_b=\bm{A}_b\bm{u}_b+\bm{\ell}_b,~~ p_b(t)\in \mathcal{S}^p_b, ~~\bm{u}_b\in \mathcal{U}_b   
\end{equation}
where $\bm{\ell}_b$ is the inflexible part of the load, $\bm{p}$ and $\bm{u}$ are column vectors, the set $\mathcal{U}_b$ is related to the flexibility that can be offered to adjust the shape of the profile, and $\mathcal{S}^p_b$ expresses operational constraints on how the asset can inject power\footnote{Later integrality constraints will be relaxed in either $\mathcal{U}_b$ or $\mathcal{S}^p_b$, and be replaced with convex constraints to guarantee a convex market clearing problem required for the convergence of the pricing algorithm.}.
Control signal constraints are mapped to $\bm{p}_b$ using $\bm{A}_b^\dagger$ as follows:
\begin{equation}
  \bm{p}_b\in \mathcal{P}_b,~~
  \mathcal{P}_b=\{\bm{p}| \bm{A}_{b}^{\dagger}(\bm{p}-\bm{\ell_{b}})\in \mathcal{U}_{b}, ~~p(t)\in \mathcal{S}^p_{b}, ~~\bm{p}^\T \bm{\lambda} \leq r_b\} \label{eq:calpb}
\end{equation}
where the constraints $\mathcal{U}_{b}$ are linear, meaning that $\bm{p}_b$ lies within a polygon. 
In \eqref{eq:calpb}, $\bm{\lambda}$ denotes the price of energy over the horizon, $r_b$ denotes the budget, and $\bm{p}_b^\T \bm{\lambda} \leq r_b$ denotes the affordability constraint.

The cost to prosumer $b$, $C_b(\bm{p}_b)$, is the price the customer is willing to pay, or the price the supplier is willing to be paid to generate for a certain amount of power. $C_b(\bm{p}_b)$ is a convex function of $\bm{p}_b$. The cost function simultaneously reflects the utility and cost for prosumers that can switch between producing and consuming, respectively, with $p_b(t)\geq 0$ representing the supply utility function and $p_b(t)< 0$ representing the demand cost function.

This generic model can be applied to various resources including renewable generation, TCLs, storage devices, DAs, and EVs. \ref{app:drmodels} shows examples of such models and cost functions for supplemental study.

\begin{figure}
    \centering
    \includegraphics[width=1\textwidth,keepaspectratio]{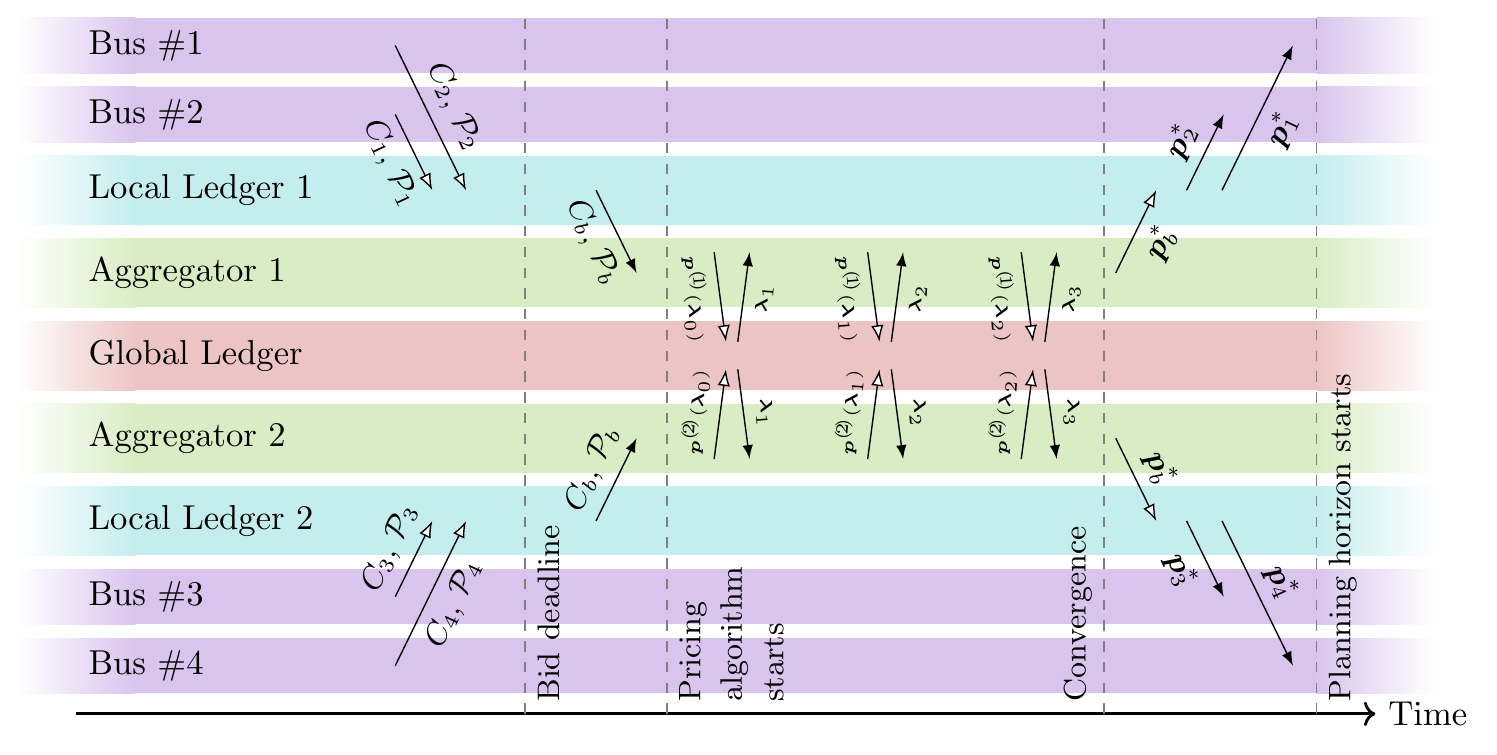}
    \caption{A single solving window for distributed pricing ordered from (i) individual buses submit their costs and constraints, (ii) aggregators collect bus costs and constraints, (iii) aggregators iterate pricing algorithm, (iv) aggregators push dispatch solution to ledger, and (v) individual buses read dispatch solution.}
    \label{fig:pricing-algorithm}
\end{figure}
\subsection{Distributed pricing and scheduling algorithm}
\label{sec.distributed}
Each aggregator can buy or sell power for the distributed resources connected to $\mathcal{B}^{(n)}$.  Let $\mathcal{B}^{(n)}=\{n(1),\dots,n(|\mathcal{B}^{(n)}|)\}$, where $n(i)$ denotes the $i$\textsuperscript{th} bus of aggregator $n$. The power profile of aggregator $n$ is: 
\begin{align}
  \bm{p}^{(n)} =\big( p^{(n)}(iT), \dots, p^{(n)}((i\!+\! 1)T\!-\!1)\big)^\T =\sum_{b\in \mathcal{B}^{(n)}}\bm{p}_b
\end{align}
where $\bm{p}_{b}$ is the profile of one of the flexible resources at a bus $b\in \mathcal{B}^{(n)}$. We shall also define the following matrix:
\begin{align}
  \bm{P}^{(n)} = \big(\bm{p}_{n(1)},\dots,\bm{p}_{n(|\mathcal{B}^{(n)}|)}\big)^\T
\end{align}
whose rows are prosumers' profiles contributing to $\bm{p}^{(n)}$. The aggregated load profile is the sum of individual components, i.e., $\bm{p}^{(n)}={\bm{P}^{(n)}}^\T \bm{1}$, where $\bm{1}=\{1\}^{\abs{\mathcal{B}^{(n)}} \times 1}$. The demand cost for aggregator $n$ for a certain set of schedules $\bm{P}^{(n)}$ is described by:
\begin{equation}
    \label{eq:demand_cost}
    \mathfrak{C}^{(n)}( \bm{P}^{(n)} )\triangleq \sum_{b \in \mathcal{B}^{(n)}}C_b(\bm{p}_b)
\end{equation}
The feasible set of matrix profiles $\bm{P}^{(n)}$ that can be chosen by the aggregator must lie in the Cartesian product of the feasible sets for the tuple of profiles $\bm{p}_b\in\mathcal{P}_b$, $b\in\mathcal{B}^{(n)}$, meaning:
\begin{align}
    \bm{P}^{(n)}\in \mathcal{P}_{n(1)}\times \dots \times \mathcal{P}_{n(|\mathcal{B}^{(n)}|)}\triangleq \mathcal{P}^{(n)}
\end{align}

\begin{algorithm}
  \caption{Prosumer pricing interaction; A step-by-step implementation of \crefrange{eq.bus-problem}{eq:convergence} from the perspective of a prosumer $b \in \mathcal{B}^{(n)}$. $LL_n$ refers to the local ledger to aggregator $n\in\mathcal{N}$, with $\Leftarrow$ and $\Rightarrow$ indicating writing to and reading from the ledger, respectively. }
  \label{alg:pricing-prosumer}
  \begin{algorithmic}[1]
    \State Wait until bidding interval starts:
    \State $LL_n \Leftarrow$ $\mathcal{P}_b$ (constraints) and $C_b$ (cost function).
    \State Wait until dispatch horizon starts.
    \If{Dispatch instruction ready on ledger}
        \State $LL_n \Rightarrow$ Read dispatch instruction $\bm{p}_b$.
        \State Execute dispatch instruction $\bm{p}_b$.
    \Else
        \State Execute most recent $\bm{p}_b$ instruction.
    \EndIf
  \end{algorithmic}
\end{algorithm}

If transmission constraints are relaxed and the algorithm only balances instantaneous power, the optimum market clearing requires solving:
\begin{align}
    \min_{\bm{P}^{(1)},\dots,\bm{P}^{(N)}} ~~ \sum_{n\in \mathcal{N}}\mathfrak{C}^{(n)}(\bm{P}^{(n)})~~
    \mbox{s.t.}~~\sum_{n\in\mathcal{N}}{\bm{P}^{(n)}}^\T \bm{1} =\bm{0},~\bm{P}^{(n)}\in \mathcal{P}^{(n)}\label{eq.market}
\end{align}
The Lagrange multiplier of the balance constraint $\sum_{n\in\mathcal{N}} {\bm{P}^{(n)}}^\T \bm{1} = 0$ in \eqref{eq.market} $\bm{\lambda}$ expresses the shadow price of energy over the horizon in \eqref{eq:calpb}, 
${\bm{\lambda}}=(\lambda(iT),\dots,\lambda((i+1)T-1))^\T$.
Also, this is an instance of the network utility maximization problems which can be decomposed as detailed next\footnote{See e.g. \cite{li2011distributed} for its application to real time pricing.}.

For each gradient descent, given the most recent price $\bm{\lambda}_{k}$ (the dual variable), each aggregator independently attempts to minimize its cost schedule through the following problem:
\begin{align}
  \label{eq.bus-problem}
  \min_{\bm{P}^{(n)}} \mathfrak{C}^{(n)}\bm{P}^{(n)}+[\bm{\lambda}_k]^\T {\bm{P}^{(n)}}^\T \bm{1}~~
  \mbox{s.t.}~~\bm{P}^{(n)}\in \mathcal{P}^{(n)},~~\forall n \in \mathcal{N}
\end{align}
Let ${\bm{P}_\star^{(n)}}(\bm{\lambda}_{k})\in \mathcal{P}^{(n)}$ be the solution of \cref{eq.bus-problem} in response to the $k$\textsuperscript{th} iteration value of the vector $\bm{\lambda}_{k}$ and  ${\bm{p}_\star^{(n)}}(\bm{\lambda}_{k})=[\bm{\lambda}_k]^\T {\bm{P}_\star^{(n)}}^\T \bm{1}$.
Assuming a feasible solution exists, the algorithm updates the price as follows in $GL$:
\begin{align}
    \label{eq.price-calc}
    \bm{\lambda}_{k+1}=\bm{\lambda}_{k}+\alpha \sum_{n\in\mathcal{N}}{\bm{p}_\star^{(n)}}^{\T}
\end{align}
Note that all the aggregators have to post their total injection based on the current price estimate, which will stop updating as soon as:
\begin{align}
    \label{eq:convergence}
    \bm{\lambda}_\star=\bm{\lambda}_{k^\star}:~~~ \sum_{n\in\mathcal{N}}{\bm{p}_\star^{(n)}}
    =\bm{0}
\end{align}
These equations comprise the distributed and decentralized algorithms explained for prosumers and aggregators in \Cref{alg:pricing-prosumer,alg:distr-pricing}, respectively, and are visualized in \Cref{fig:pricing-algorithm}. 

\begin{algorithm}
  \caption{Aggregator pricing algorithm; A step-by-step implementation of \crefrange{eq.bus-problem}{eq:convergence} from the perspective of aggregator $n \in \mathcal{N}$, with $\Leftarrow$ and $\Rightarrow$ indicating writing to and reading from the ledger, respectively. }
  \label{alg:distr-pricing}
  \begin{algorithmic}[1]
    \State Define algorithm time limit $\overline{\tau}$ and iteration limit $\overline{k}$.
    \State Wait until prosumer bidding interval ends.
    \State $LL_n \Rightarrow$ read all $C_b$, $r_b$, and $\mathcal{P}_b$ values available on ledger, building $\mathcal{B}^{(n)}$ based on submitted prosumers.
    \State Initialize $\bm{\lambda}_\triangle$ and $\bm{p}_\triangle^{(n)}$ to most recent solutions $\bm{\lambda}_\star$ and ${\bm{p}_\star^{(n)}}$, or $\bm{0}$ if no prior solution exists.
    \State Initialize $\bm{\lambda}_0 \gets \bm{\lambda}_\triangle$, $k\gets 0$ and $\hat{\alpha}$.
    \State Build model \cref{eq.bus-problem} using $\bm{\lambda}_0$.
    \While{$k < \overline{k}$}
      \State Solve model \cref{eq.bus-problem} for $\bm{\lambda}_k$, enforcing the billing constraint for all prosumers.
    \State Retry writing solution ${\bm{p}_\star^{(n)}}$ to $GL$ until ACK received.
      \State Start timer $\tau \gets 0$. 
      \While{Other aggregator solutions are not available from $GL$} 
        \If {$\tau < \overline{\tau}$}
          \State Wait $\varepsilon$ seconds.
        \Else
          \State $LL_n \Leftarrow$ $\bm{p}_b^\triangle$ for all prosumers $b\in\mathcal{B}^{(n)}$.
          \State $LL_n \Leftarrow$ $r_b \gets r_b +  {\bm{p}_b^\triangle}^\T\bm{\lambda}_\triangle$ for all prosumers $b\in\mathcal{B}^{(n)}$.
          \State Terminate algorithm, recycling last solutions.
        \EndIf
      \EndWhile
      \State $GL \Rightarrow {\bm{p}_{\star}^{(m)}}^\T \bm{\lambda}_k$ for all $m\in\mathcal{N} \setminus \{n\}$. 
      \State Update $\bm{\lambda}_{k+1}$ following \cref{eq.price-calc} using $\alpha = \hat{\alpha}/(k+1)$. 
      \If{$\abs{\bm{\lambda}_{k+1} - \bm{\lambda}_k} < \epsilon$} \label{step:ep_alg_2}
        \State $\bm{\lambda}_\star \gets \bm{\lambda}_k$.
        \State $LL_n \Leftarrow$ $\bm{p}_b^\star$ for all prosumers $b\in\mathcal{B}^{(n)}$.
        \State $LL_n \Leftarrow$ $r_b \gets r_b + {\bm{p}_b^\star}^\T \bm{\lambda}_\star$ for all prosumers $b\in\mathcal{B}^{(n)}$.
        \State Terminate.
      \EndIf
      \State $k \gets k + 1$.
    \EndWhile
    \State Terminate.
  \end{algorithmic}
\end{algorithm}
A few interesting observations can be made: 
\begin{enumerate}[leftmargin=*]
  \item Aggregators hide local bids and feasibility constraints from the $GL$, and instead they only show (expose) how they would react for the particular price scenarios iterated through $\bm{\lambda}_k$. 
  \item Because the cost and constraints of an aggregator are decomposable in terms of the prosumer profiles $\bm{p}_b$, the problem could be decomposed further to allow individual prosumers to keep $C_b(\bm{p}_b)$ and $\mathcal{P}_b$ private, interacting with the aggregator in a similar manner as shown in~\cite{karakoc2018multi}.  
\end{enumerate}

There are challenges with convergence of the distributed algorithm \cref{eq.bus-problem} not commonly found in comparable centralized formulations. First, the aggregate supply/demand curves must cross for a fixed price to emerge, a standard requirement in market theory. Second, a rogue aggregator may produce malicious values of $p_\star^n$ during the iterative phase of \Cref{alg:distr-pricing}, potentially preventing the algorithm from converging. There are numerous ways to detect such manipulations such as bounding the gradient step \cref{eq.price-calc}, however, the details of those are beyond the scope of this paper and readers are referred to~\cite{ravi2019detection} for additional information. Third, due to the non-convexity introduced by the budget constraint in \cref{eq:calpb}, there are no guarantees for convergence to the global optimum. However, as this constraint is only tight for a small number of prosumers, the problem often converges in practice. 

\section{EVE Distributed Robust State Verification Algorithm}
\label{sec:verification}
While the blockchain TE framework ensures transparency and immutability of transactions placed on the chain, it is still insufficient to ensure that those transactions took place. Physical measurements are needed to confirm power injections at each bus in the electric grid. In a trustless system, aggregators have to cooperate to verify measurement accuracy, noting that some aggregators (or some of the prosumers they manage) may operate as adversaries by modifying market operations and/or measurements to commit energy theft. The essential tool described in \Cref{sec:physical} establishes accuracy by checking consistency with the grid's physical laws.

Before delving into algorithmic details, it is worth emphasizing that the notion of \textit{distributed measurement verification} introduced here is new. The approach proposed for measurement verification consists of solving a regression problem closely related to state estimation in power systems fitting sensor data, but its goal is fundamentally different. Securely estimating the entire state vector for the grid is sufficient but not necessary to cross-validate the self-reported measurements. The state variables themselves are therefore not essential, and rather, we focus on the accuracy of reported \textit{real power injections} within an aggregator region and \textit{power flows between} different aggregators' regions. This approach ensures appropriate billing aggregators and prosumers and permits penalties to be levied for discrepancies in reporting. For measurements of the consumption/generation in time window $\mathcal{T}_{i}$, we run the verification mechanism in time window $\mathcal{T}_{i+1}$ as shown in \Cref{fig:timeline}. Considering an FDIA threat \cite{liu2011false}, our mechanism extends the idea in~\cite{vukovic2014security} and adapts it to serve as a decentralized cross-validation algorithm integrated within the EVE framework. The goal is to extrapolate the actual power injections $\bm{p}^{(n)}$ of each aggregator $n\in \mathcal{N}$ from sensor measurements. Since $\bm{p}^{(n)}$ are continuous variables, this is not a binary decision and amounts to solving a regression problem. 

\subsection{Physical Constraints for the Electric Grid}
\label{sec:physical}
\textcolor{black}{Let $\bm{x}(t)$ be the vector of system variables at time $t$, consisting of bus injection variables and branch flow variables. Here, the variables in $\bm x(t)$ include real power injection, reactive power injection, and squared voltage magnitude expressed as $(p_b(t),q_b(t), v_b^2(t))^{\T},~\forall b \in \mathcal{B}$.} Branch flow variables include the squared current magnitudes, real power flows, and reactive power flows expressed as $(c_l^2(t), P_l(t), Q_l(t))^{\T},~\forall l \in \mathcal{E}_e$. Let $\bm{x}_{\mathcal{A}}(t)$ include only the variables in $\bm x(t)$ that are directly-measured through a sensor. Measurements are noisy versions of physical parameters described by: 
\begin{equation}
    \label{eq.sensors}
    \bm{z}(t) = \bm{x}_{\mathcal{A}}(t) + \epsilon(t). 
\end{equation}
Within a margin of error due to the noise, these physical constraints are a set of non-linear homogeneous equations written in vector form as follows:
\begin{align}
    \label{eq:elec_grid_const}
    \bm{h}(\bm{x}(t)) = 0~; ~~ \forall t \in \mathcal{T}
\end{align}
\ref{app:StateVer} specifies a possible  $\bm{h}(\cdot)$ using the Distflow \cite{barenwunetworkreconfig} equations .

\subsection{Malicious Agents Behavior}
\label{sec:malbeh}
We assume that the adversary (a malicious agent or a group of coordinating agents) is an insider who has legitimate physical and logical access to the network and ledgers through the certification mechanism. We also assume that the adversary is capable of manipulating sensors measurements, either by compromising sensors or compromising the communication between sensors and aggregators. The insider is motivated to disrupt the verification process and cheat the system for financial gain.

Utilizing sensor measurements reported at the $LL$ level in \cref{eq.sensors} and physical constraints in \cref{eq:elec_grid_const}, we formulate a decentralized optimization algorithm to complete the verification task using data from all sensors under any aggregator. The optimization algorithm (detailed later in \Cref{sec:RSVProblem}) may be generalized into the following form:
\begin{subequations}\label{eq:decomp-opt_prob}
\begin{align}
    \min_{\bm{x}^{(n)}} &~~ \sum_{n\in\mathcal{N}} f^{(n)}(\bm{x}^{(n)})\label{eq:decomp-opt}\\
    \text{s.t.} &~~ \text{Consensus Constraints}\label{eq:decomp-opt-constraints}
\end{align}
\end{subequations}
where $n$ is a decentralized agent in set $\mathcal{N}$\footnote{We abuse the notation for the set of aggregators $\mathcal{N}$ to denote the set of decentralized agents in this subsection.} and $f^{(n)}:\mathbb{R}^{l}\rightarrow \mathbb{R}$ is a cost function that agent $n$ has to minimize while cooperatively minimizing the aggregate of cost functions from all the agents \cref{eq:decomp-opt}, subject to some consensus constraints \cref{eq:decomp-opt-constraints}. Common algorithms used to solve this problem using a certain communication graph $\mathcal{G}_c(\mathcal{N},\mathcal{E}_c)$ \cref{eq:decomp-opt_prob} involve iterative consensus updates to $\bm{x}^{(n)} \in \mathbb{R}^{l}$ as follows:
\begin{align}
    \label{eq:consensusex}
    \bm{x}^{(n)}_{k+1} = \sum_{m\in\mathcal{N}} a_{nm} \bm{r}^{(m)}(\bm{x}^{(m)}_k) ~~\textnormal{at}~~k \geq 0
\end{align}
where $k$ is the iteration index, $a_{nm} \geq 0$ is a mixing weight  $(n,m)\in \mathcal{E}$ such that $\sum_{m\in \mathcal{N}}a_{nm} = 1$, and $\bm{r}^{(m)}:\mathbb{R}^{l}\rightarrow \mathbb{R}^{l}$ is the information received by $n$ from neighbor $m$. We see from \cref{eq:consensusex} that neighbors on the communication graph exchange information with one another in each iteration.

The communications model is described as:
\begin{equation}
\label{eq:falseinject}
\bm{r}^{(m)}(\bm{x}^{(m)}_k) =  \begin{cases}
                        \bm{x}^{(m)}_k, & m\in\mathcal{R}\\
                        \star, & m\in\mathcal{M}
                    \end{cases}
\end{equation}
with $\mathcal{R}\subseteq \mathcal{N}$ and $\mathcal{M}\subseteq \mathcal{N}$ expressing the sets of regular and malicious agents, respectively. That is, regular agents report their true states, whereas malicious agents may inject false data and disrupt convergence of the algorithm to suit their goals.

An attack by an agent results in the following update for a neighbor $n$:
\begin{align}
    \label{eq:modifiedconsensus}
    \widetilde{\bm{x}}_{k+1}^{(n)} = \bm{x}_{k+1}^{(n)} + \star
\end{align}
where $\widetilde{\bm{x}}_{k+1}^{(n)}$ is the false update and $\bm{x}_{k+1}^{(n)}$ is the true update (if all the neighbors communicated truthfully). For instance, a malicious agent will attempt theft by paying less (as a consumer) or receiving a larger compensation (as a producer). This occurs by under reporting energy usage or over reporting generation by altering the appropriate field of $\bm{z}^{(n)}(t)$ so that:
    $\widetilde{\bm{z}}^{(n)}(t) = \bm{z}^{(n)}(t) + \star$.
\subsection{{Robust State Verification in the Presence of FDIAs}}
\label{sec:state_verification}
Next, we provide details on the decentralized algorithm that allows aggregators to cross verify if measurements of power injections are to be trusted and, if not, which aggregators are likely responsible for the FDIA. 

{
\subsubsection{Modelling of the Optimization Problem}\label{sec:RSVProblem}
We pose the state verification problem as a decentralized optimization problem in which each aggregator $n \in \mathcal{N}$ has access only to their private cost function $f^{(n)}: \mathbb{R}^{l_n} \rightarrow \mathbb{R}$ and act on their own private vector of system variables, $\bm{x}^{(n)}(t)$. Here, $\bm{x}^{(n)}(t)$ includes copies of those variables in $\bm{x}(t)$ that pertain to buses and lines inside and at the periphery of aggregator region $n$ (see \Cref{fig:state_est}). That is,
\begin{align}
    \bm{x}^{(n)}(t) = \bm{S}^{(n)}\bm{x}(t)
\end{align}
where $\bm{S}^{(n)}$ is a selection matrix that extracts the appropriate entries that make up $\bm{x}^{(n)}(t)$ from $\bm{x}(t)$.
Since there are tie-lines between aggregator regions, some of the entries in $\bm{x}^{(n)}(t)$ will have identical counterparts in $\bm{x}^{(m)}(t)$ of a neighboring aggregator region $m$. 
For the extrapolated injections to be valid, the regions have to match values at these tie-lines. Failure to do so indicates an attack. The consensus constraint is, therefore, a consensus on the tie-lines variables, that can
be written as:
\begin{align}\label{eq:cons_constr1}
    \bm{S}_{nm}\bm{x}^{(n)}(t) = \bm{S}_{mn}\bm{x}^{(m)}(t), \qquad \forall n \in \mathcal{N} \text{ and } m : nm \in \mathcal{E}_c
\end{align}
where $\bm{S}_{nm}$ is the selection matrix that extracts from $\bm{x}^{(n)}$ common variables between neighboring regions $n$ and $m$. 

Thus, the goal of the set of aggregators is to minimize their individual cost functions ($f^{(n)}, \forall n \in \mathcal{N}$) and the global cost function ($\sum_{n\in \mathcal{N}} f^{(n)}$) simultaneously subject to the consensus constraints \cref{eq:cons_constr1} defined over the communication graph, $\mathcal{G}_c(\mathcal{N},\mathcal{E}_c)$.

The optimization problem should find values of system variables, $\bm{x}^{(n)}(t)$, at each aggregator $n$ that:
\begin{itemize}
    \item Have the least residual error with respect to the measurements, $\bm{z}^{(n)}(t)$, to reduce measurement deviation from their corresponding system variables ($\bm{x}_{\mathcal{A}}^{(n)}(t)$) being minimized. Letting $\bm{S}_{\mathcal{A}}^{(n)}$ be the selection matrix that extracts available measurements from $\bm{x}^{(n)}(t)$, we can write
    \begin{align}
        \bm{x}_{\mathcal{A}}^{(n)}(t) = \bm{S}_{\mathcal{A}}^{(n)} \bm{x}^{(n)}(t)
    \end{align}
    \item Have the least residual error with respect to the scheduled injections, ${\bm{\mathfrak{p}}^\star}^{(n)}(t)$, where the $i$\textsuperscript{th} element of the vector is given by $[{\bm{\mathfrak{p}}^\star}^{(n)}(t)]_i = p_{b_i}(t),~\forall b_i \in \mathcal{B}^{(n)}$. This seeks to minimize deviation of the scheduled bus dispatch from their corresponding system variables ($\bm{S}_{\mathcal{P}}^{(n)}\bm{x}^{(n)}(t)$), where $\bm{S}_{\mathcal{P}}^{(n)}$ is the selection matrix that extracts active power injections variables from $\bm{x}^{(n)}(t)$, and
    \item Fit the physical model equations, $\bm{h}^{(n)}(\cdot)$\footnote{Definitions of the functions $\bm{h}^{(n)}(\cdot)$ are given in \ref{app:StateVer}.}, with the least residual error. 
\end{itemize}
Power injection components, $\bm{S}^{(n)}_{\mathcal{P}}\bm{x}^{(n)}$, from the optimization problem are used in EVE as the closest approximation for ground-truth billing. The severity of deviations from the schedule $\left({\mathfrak{p}^\star}^{(n)}(t)-\bm{S}^{(n)}_{\mathcal{P}}\bm{x}^{(n)}\right)$ that determines the penalties assigned to prosumers under a specific bus. 

Finally, the state verification problem can be cast as the following optimization problem, written in a form analogous to \cref{eq:decomp-opt} that is amenable to ADMM decomposition~\cite{boyd2011distributed}:
\begin{subequations}\label{eq:RSVproblem}
\begin{align}
    \min_{\{\bm{x}^{(n)}(t)\}^{t\in\mathcal{T}}_{n\in \mathcal{N}}}~&\!\!\!
    \sum_{t\in\mathcal{T}}
    \sum_{i\in\mathcal{N}} f^{(n)}(\bm{x}^{(n)}(t))
    \label{eq:State_Est}
       \\
    \text{s.t.} ~&~ \bm{S}_{nm}\bm{x}^{(n)}(t) = \bm{S}_{mn}\bm{x}^{(m)}(t)~,~~\forall nm \in \mathcal{E}_c\label{eq.consensus}\\
    ~&~ \bm{x}^{(n)}(t)=\bm{S}^{(n)}\bm{x}(t)~,~~\forall n\in\mathcal{N}\\
    ~&~
    \bm{x}_{\mathcal{A}}^{(n)}(t)=\bm{S}_{\mathcal{A}}^{(n)}\bm{x}(t), ~~\forall n\in\mathcal{N}
\end{align}
\end{subequations}
where:
\begin{multline}
    f^{(n)}(\bm{x}^{(n)}(t)) := c_1 \norm{\bm{\Sigma}_{z^{(n)}}^{-1/2} \pqty{\bm{z}^{(n)}(t) - \bm{x}^{(n)}_{\mathcal{A}}(t)}}_2^2 + c_2 \norm{\bm{h}^{(n)}(\bm{x}^{(n)}(t) )}^2_2 + \\ c_3 \norm{ {\bm{\mathfrak{p}}^\star}^{(n)}(t) - \bm{S}^{(n)}_{\mathcal{P}}\bm{x}^{(n)}(t)}^2_2
\end{multline}
Here, $\bm{\Sigma}_{z^{(n)}}$ is the diagonal matrix such that $[\bm{\Sigma}_{z^{(n)}}]_{jj} = \textnormal{variance}([\bm{z}^{(n)}]_j)$. Additionally, \cref{eq.consensus} enforces consensus among the common variables across neighboring regions for the constraints mentioned in \cref{eq:decomp-opt-constraints}.

\begin{remark}
The terms $c_1, c_2, c_3$ denote the weight of the penalty levied on the solution if their respective objectives are violated. The third term penalizes a solution that is further from the schedule, and in noting this might hinder the verification step, practical applications can choose $c_3$ such that $c_3 \lll c_1, c_2$. 
\end{remark} 
\begin{figure}
\centering
    \includegraphics[scale=0.65,keepaspectratio]{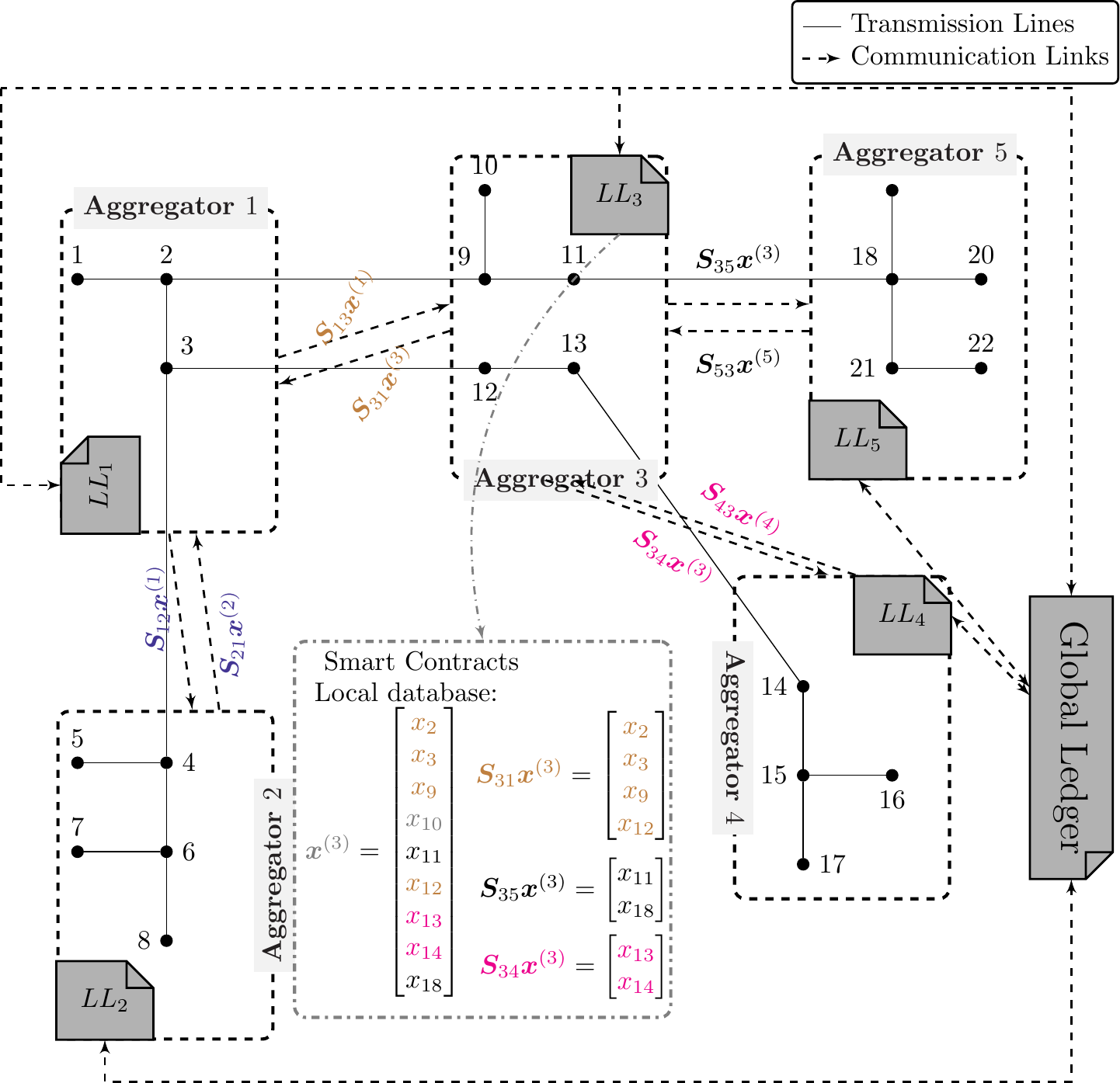}
    \caption{State Verification Architecture for a test case with five aggregator regions.}
\label{fig:state_est}
\end{figure}
}

\subsubsection{Iterative Update Rule}
In this section, we present equations that iteratively update state variables vectors to arrive at the solution to problem \cref{eq:RSVproblem}. First, we define two matrices that assist in formulating rules for iterative updates. Suppose that $l_n$ is the number of variables pertaining to aggregator region $n$, then $\bm{D}^{(n)} \in \mathbb{R}^{l_n \times l_n}$ is a diagonal matrix with $[\bm{D}^{(n)}]_{jj}$ equal to the number of regions with which region $n$ has the $j$-th variable of $\bm{x}^{(n)}$ in common, $\forall j \in [l_n]$. Similarly $\overline{\bm{D}}^{(n)}\in \mathbb{R}^{l_n \times l_n}$ is also a diagonal matrix such that $[\overline{\bm{D}}^{(n)}]_{jj} =1/[\bm{D}^{(n)}]_{jj}$ if $[\bm{D}^{(n)}]_{jj}\neq 0$, and $[\overline{\bm{D}}^{(n)}]_{jj} =0$ if $[\bm{D}^{(n)}]_{jj} = 0$.

To arrive at the minimizer of the problem in \cref{eq:RSVproblem} for an interval $t \in \mathcal{T}_i$, the updates in \cref{eq:ADMM_updates} are executed iteratively $\forall k\geq 0$ until a termination condition is satisfied\footnote{For the sake of readability, we write $\bm{x}^{(n)}$ instead of $\bm{x}^{(n)}(t)$ and use $\bm{x}^{(n)}_k$ to denote iterates of the algorithm where $k$ is the iterate.}:
\begin{subequations}\label{eq:ADMM_updates}
\begin{align}
    \bm{x}^{(n)}_{k+1} =& \left(c_1{\bm{S}^{(n)}_{\mathcal{A}}}^\T{\bm{\Sigma}}_{z^{(n)}}^{-1} \bm{S}^{(n)}_{\mathcal{A}} + c_2{\bm{H}^{(n)}}^\T \bm{H}^{(n)}   + c_3{\bm{S}^{(n)}_{\mathcal{P}}}^\T\bm{S}^{(n)}_{\mathcal{P}} + c_4\bm{D}^{(n)}\right)^{-1}\nonumber\\
    & \times \left( c_1{\bm{S}^{(n)}_{\mathcal{A}}}^\T{\bm{\Sigma}}_{z^{(n)}}^{-1}\bm{z}^{(n)} + c_3{\bm{S}^{(n)}_{\mathcal{P}}}^\T{\bm{\mathfrak{p}}^\star}^{(n)} + c_4\bm{D}^{(n)}\bm{\upsilon}^{(n)}_k \right)\label{eq:ADMM_x}\\
    \bm{\psi}^{(n)}_{k+1} =& \overline{\bm{D}}^{(n)} \sum_{m: nm \in \mathcal{E}_c} \bm{S}_{nm}^\T\bm{S}_{mn}\bm{x}^{(m)}_{k+1}\label{eq:ADMM_phi}\\
    \bm{\upsilon}^{(n)}_{k+1} =& \bm{\upsilon}^{(n)}_{k} + \bm{\psi}^{(n)}_{k+1} - 0.5\left(\bm{\psi}^{(n)}_{k} + \bm{x}^{(n)}_{k} \right)\label{eq:ADMM_ups}
\end{align}
\end{subequations}
where $\bm{H}^{(n)}$ is defined in \ref{app:StateVer} with variables initialized as:
\begin{subequations}\label{eq:ADMM_inits}
\begin{align}
    \bm{x}^{(n)}_0 &\in \mathbb{R}^{l_n}\qquad\ldots\text{initialize arbitrarily}\\
    \bm{\psi}^{(n)}_0 &= \overline{\bm{D}}^{(n)} \sum_{m: nm \in \mathcal{E}_c} \bm{S}_{nm}^\T \bm{S}_{mn} \bm{x}^{(m)}_{0}\\
    \bm{\upsilon}^{(n)}_0 &= \frac{1}{2}\left(\bm{\psi}^{(n)}_{0} + \bm{x}^{(n)}_{0} \right).
\end{align}
\end{subequations}

\Cref{alg:state_verification} describes the iterative process in \cref{eq:ADMM_updates} with a pictorial representation of the algorithm given in \Cref{fig:state_est}. Buses in each aggregator region $n$ have access to their local ledgers $LL_n$. The aggregators, in addition to their local ledger, also have access to $GL$. The aggregator collects measurements from their buses and stores them on $LL_n$. Results after an ADMM update are also saved in $LL_n$ for each aggregator. Aggregators exchange state elements that correspond to tie-line variables with a neighbor as shown in \Cref{fig:state_est}. 

\begin{algorithm}
\caption{Robust state verification; A step-by-step implementation from the perspective of aggregator $n~\forall n \in \mathcal{N}$. Here $LL_n~\forall n \in \mathcal{N}$ and $GL$ represents the local and global ledgers respectively. The symbol $a \Leftarrow b$ corresponds to upload from $b$ to $a$.}
\label{alg:state_verification}
\begin{algorithmic}[1]
\State $LL_n \Leftarrow$ \text{ Collect local measurements} $\bm{z}^{(n)}(t)$.
\State \label{algstep:Init}Initialize ADMM states according to \cref{eq:ADMM_inits}, trust score $\bm{\pi}= \bm{0}$, and disagreements $d_{nm} = 0,~\forall nm \in  \mathcal{E}_c$.
\Repeat
    \State \label{step:pi_init}$\bm{\pi}^{-} \gets \bm{\pi}$ and $[\bm{x}^{(n)}(t)]^{-}\gets \bm{x}^{(n)}(t)$.
    \State ADMM update of $\bm{x}^{(n)}(t)$ according to equation~\cref{eq:ADMM_x}.
        \State \parbox[t]{\dimexpr\textwidth-\leftmargin-\labelsep-\labelwidth}{Send common variable information to neighbors via the $GL$ with ABAC control: $GL \Leftarrow \{\bm{S}_{ij}\bm{x}^{(n)}(t) \mid m: nm\in \mathcal{E}_c\}$. \strut}
        \State \parbox[t]{\dimexpr\textwidth-\leftmargin-\labelsep-\labelwidth}{Receive common variable information from neighbors:\\$\{\bm{S}_{mn}\bm{x}^{(m)}(t) \mid m : nm\!\in\!\mathcal{E}_c \text{, $m$ sent information}\}\bigcup\\\{\bm{S}_{mn}\bm{x}^{(m)}(t-1)\mid m:nm\in\mathcal{E}_c, m \text{ didn't send}  \text{information}\}~\Leftarrow~GL$.\strut}
        \State Update intermediate states according to \cref{eq:ADMM_phi,eq:ADMM_ups}.
        \State Run \Cref{alg:Security} to update $\bm{\pi}$ and $d_{nm},~\forall m: nm\in \mathcal{E}_c$\label{algstep:subroutine}.
\Until {One of the termination conditions \ref{term:conv} or \ref{term:attac} is satisfied}
\State Restart the algorithm with $\{\mathcal{G}_c\setminus  \left(\arg\max_n {\pi_n}\right)\}$.\label{algstep:restartverif}
\end{algorithmic}
\end{algorithm}

\subsubsection{Threat Model specific to RSV}
\label{sec:threat_model}

The updates in \crefrange{eq:ADMM_phi}{eq:ADMM_ups} involve aggregating the shared state variable $\bm{x}^{(n)}$ from agent $n$  with the same state variable from its neighbors, similar to the updates in \cref{eq:consensusex}. Yet \Cref{sec:malbeh} showed that approach to be vulnerable to FDI attacks via \cref{eq:modifiedconsensus}, thus indicating \cref{eq:ADMM_phi} is also vulnerable to malicious injections by neighbors.

A malicious user in region $m$ may modify values of its input measurements to affect the aggregator's measurement vector as follows:
\begin{align*}
    \widetilde{\bm{z}}^{(m)} = \bm{z}^{(m)} + \bm{a}^{(m)}
\end{align*}
where the perturbation $\bm{a}^{(m)}$ has non-zero entries in the locations that correspond to false sensor measurements.
Similarly, if the aggregator itself acts maliciously, it can inject false data (as seen in \cref{eq:falseinject}) into the updates of $\bm{S}_{mn}\bm{x}^{(m)}$ that are passed to a neighbor aggregator $n$ in \cref{eq:ADMM_phi}. Both changes lead to discrepancies in neighboring aggregators updates as follows:
\begin{align}
    \widetilde{\bm{\psi}}^{(n)}_{k}(t) =& {\bm{\psi}}^{(n)}_{k}(t) + \overline{\bm{D}}^{(n)}\bm{S}_{nm}^\T\bm{S}_{mn}\bm{a}^{(m)}_{k}(t)\label{eq:modified_ADMM_phi}\\
    \Rightarrow~~\widetilde{\bm{\upsilon}}^{(n)}_{k}(t) =& {\bm{\upsilon}}^{(n)}_{k}(t) + \overline{\bm{D}}^{(n)}\bm{S}_{nm}^\T\bm{S}_{mn}\bm{a}^{(m)}_{k}(t)\label{eq:modified_ADMM_ups}\\
    \widetilde{\bm{x}}^{(n)}_{k+1}(t) =& {\bm{x}}^{(n)}_{k+1}(t) + c_4\bm{M}\bm{D}^{(n)}\overline{\bm{D}}^{(n)}\bm{S}_{nm}^\T\bm{S}_{mn}\bm{a}^{(m)}_{k}(t)\label{eq:modified_ADMM_x}
\end{align}
where $\bm{M}\!\!=\!\!\left(\!\!{\bm{H}^{(n)}}^\T\bm{H}^{(n)} + {\bm{S}^{(n)}_{\mathcal{A}}}^\T\bm{S}^{(n)}_{\mathcal{A}} + {\bm{S}^{(n)}_{\mathcal{P}}}^\T\bm{S}^{(n)}_{\mathcal{P}} + c_4\bm{D}^{(n)}\right)^{-1}$ and \cref{eq:modified_ADMM_x} is analogous to the discrepancy in \cref{eq:modifiedconsensus} where the inaccurate update is a modified version ($\widetilde{\bm{x}}^{(n)}_{k+1}(t)$) of the true update ($\bm{x}^{(n)}_{k+1}(t)$).

The FDIAs will be successful at creating algorithm divergence, or convergence to a false optimum, if the ADMM updates in \crefrange{eq:ADMM_x}{eq:ADMM_ups} are used to solve \cref{eq:State_Est}. Algorithm divergence is a special case of a \textit{Denial of Service} attack in which aggregators are unable to complete the verification process. 

An additional area of concern is stealth attacks where the attacker injects a sparse vector, $\bm{a}^{(n)}$. Here, non-zero entries of the attack vector correspond to the sensors being attacked, such that the constraint in \cref{eq:stealth_attack} is satisfied even with the perturbed state: 
\begin{align}
    \label{eq:stealth_attack}
    \bm{h}^{(n)}(\bm{x}^{(n)} + {\bm{S}_{\mathcal{A}}^{(n)}}^\T \bm{a}^{(n)}) = 0
\end{align}
where $\bm{x}^{(n)}$ corresponds to the true variables. Here, without any change in the loss function in the state verification problem \cref{eq:State_Est}, the attacker is still able to alter the algorithm's output. These types of attacks are only possible when a malicious aggregator can gain complete knowledge about its neighbors' parameters. Such attacks are tough to detect, and even harder to mitigate, in the absence of a specially imposed structure on the actual measurement vectors. In practice, specially designed sparsity patterns for sensors can prevent such attacks.

\subsubsection{Detection of the Malicious Agent}\label{sec:attac_detect}
Methods proposed in \cite{vukovic2014security} are employed to detect an attack as presented in \Cref{alg:Security}. \Cref{alg:Security} is a detection subroutine with the robust state verification of \Cref{alg:state_verification}.
\begin{algorithm}
\caption{Detection loop; $F(d_{nm}, \bm{S}_{nm}\bm{x}^{(n)}(t), \bm{S}_{ji}\bm{x}^{(m)},~\forall m: nm\in \mathcal{E}_c$)}
\label{alg:Security}
\begin{algorithmic}[1]
\State Set $\alpha$, $\epsilon=10^{-16}$
\State Calculate $d_{nm}~\forall m: nm \in \mathcal{E}_c$ according to \cref{eq:dij}.
\State Calculate $[\bm{B}]_{nm} \forall m: nm \in \mathcal{E}_c$ according to \cref{eq:Bcalc}.
\State Submit $[\bm{B}]_{n:}$ to $GL$ until ACK received.
\State $[\bm{B}]_{m:} \Leftarrow GL, \forall m \in \mathcal{N}\setminus\{n\}$
\State \label{algstep:pi_calc}Compute $\bm{\pi}$, the left principle eigenvector of $\bm{B}$.
\State \Return $\bm{\pi}$ and $d_{nm},~\forall m: nm\in \mathcal{E}_c$.
\end{algorithmic}
\end{algorithm}

In \Cref{alg:Security}, each region $n$ calculates a measure of disagreement, $d_{nm}$, in the shared variables with a neighboring region $m$ as:
\begin{align}
\label{eq:dij}
    d_{nm}  &= (1-\alpha_k)d_{nm} 
    +\frac{\alpha_k/4}{|\bm{S}_{nm}\bm{x}^{(n)}(t)| |\mathcal{T}_i|} \sum_{t\in\mathcal{T}_i}{\norm{\bm{S}_{nm}\bm{x}^{(n)}(t)\!-\!\bm{S}_{mn}\bm{x}^{(m)}(t)}_F^2} 
\end{align}
and the matrix of normalized disagreement scores $\bm{B}$:
\begin{align}
    \label{eq:Bcalc}
    [\bm{B}]_{nm} &= \dfrac{d_{nm}}{\sum_{m': nm' \in \mathcal{E}_c} d_{nm'}+\epsilon}.
\end{align}
 The left principal eigenvector, $\bm{\pi}$, of $\bm{B}$ is then calculated. The value of $\norm{\bm{\pi}}_2$ and the location of the highest element of $\bm{\pi}$ represent the presence of an attack and the index of the most likely attacker, respectively \cite{vukovic2014security}.  

To mitigate the impact of FDIA, \cref{algstep:restartverif} is added to \Cref{alg:state_verification} to restart the algorithm after isolating the identified attacker. However, the structure of the communication graph can cause misidentification errors, resulting in divergence in the proposed algorithm. Convergence can be guaranteed provided the following condition is met when establishing the structure of the communication graph: 
\begin{theorem}[Proposition 3~\cite{vukovic2014security}]\label{thm:3-cliques}
Consider a system with $N>2$ regions, if $(i)$ there exists a $3$-clique in the graph $\mathcal{G}_c$ and if $(ii)$ for finite $k$ the  RSV does not converge, then the stationary distribution $\bm{\pi}_k$ exists and it is unique and can be computed.
\end{theorem}
\subsubsection{Termination Conditions}
\Cref{alg:state_verification} terminates when either of the following two conditions is satisfied. 
\begin{enumerate}[label= T\arabic*.,leftmargin=*]
    \item\label{term:conv} The first condition is met when an aggregator converges, i.e., 
    \begin{equation}
        \label{eq:t1}
        \norm{\bm{x}^{(n)}_k(t) - \bm{x}^{(n)}_{k-1}(t)}_\infty \leq \epsilon
    \end{equation} 
    \item\label{term:attac} The second condition is met when all aggregators have agreed about the presence and identity of an FDI attacker, i.e.,
    \begin{equation}
        \label{eq:t2}
        \norm{\bm{\pi}_k - \bm{\pi}_{k-1}}_\infty \leq \epsilon_\pi \quad\text{and}\quad \pi_m > \mu_m(\bm{\pi}) + \beta\sigma_m(\bm{\pi})
    \end{equation}
     for some $\beta>0,\,m\in \mathcal{N}_n$, where each agent $n$ calculates $\mu_m(\bm{\pi})$ and $\sigma_m(\bm{\pi})$ as the excluded average and excluded standard deviation, respectively:
    \begin{equation}
    \begin{aligned}
        {\mu}_m(\bm{\pi}) &= \frac{1}{|\mathcal{N}_n|-1}\textstyle{\sum_{m'\in\mathcal{N}_n\setminus\{m\}}} \pi_{m'} \\
        \sigma_m(\bm{\pi}) &= \sqrt{\frac{1}{|\mathcal{N}_n|-1}\sum_{m'\in\mathcal{N}_n\setminus\{m\}} |\pi_{m'}-\mu_{m'}(\bm{\pi})|^2}
    \end{aligned}
    \end{equation}
\end{enumerate}

\subsubsection{Placement of Measuring Instruments}
\label{sec:placement_measurement}
Partitioning the electric grid into $N$ aggregator regions must be done according to a ruleset. Recovering a unique $\bm{\pi}_k$ during the convergence failure of the RSV algorithm requires the presence of at least one $3$-clique in the communication graph $\mathcal{G}_c$ (Theorem~\ref{thm:3-cliques}). However, the radial structure of the distribution grid and the requirement for each aggregator region to contain a contiguous set of buses limits the number of possible $3$-cliques on $\mathcal{G}_c$. A possible solution is to allow aggregators to access a small amount of sensor measurements from the neighboring aggregators to improve accountability at the expense of privacy. 


\section{Numerical Simulation}
\label{sec:simulations}
This section demonstrates performance of the distributed pricing and robust state verification algorithms before describing implementation on HLF in \Cref{sec:imp}.  
\subsection{Simulation Setup}
\begin{figure}[tbp]
    \centering
    \includegraphics[width=\linewidth]{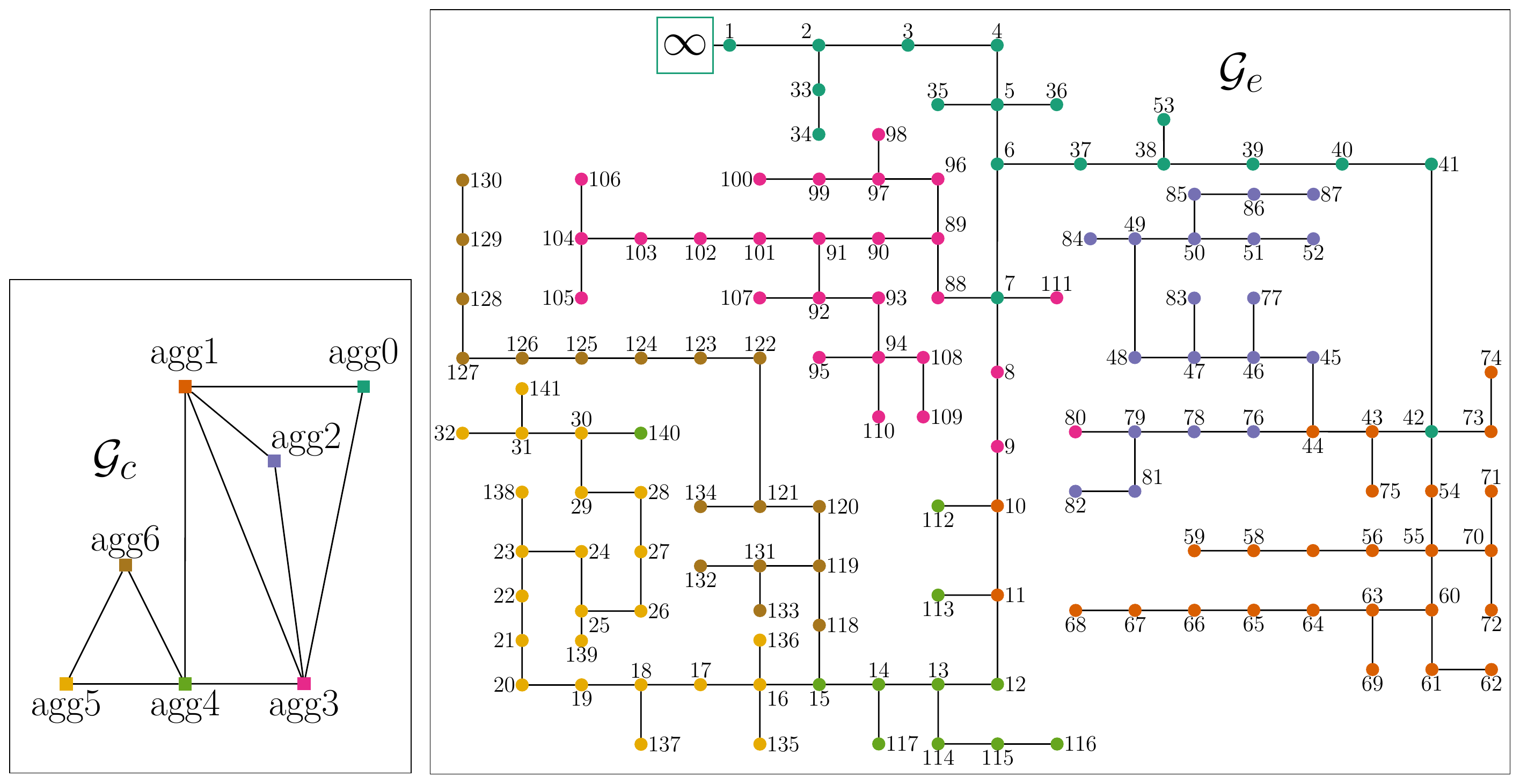}
    \caption{$\mathcal{G}_e$ is the network graph corresponding to the $141$ bus radial distribution network. The network in the box shows the communication graph $\mathcal{G}_c$ with the nodes representing aggregators.}
    \label{fig:case141}
\end{figure}

Figure \ref{fig:case141} shows the radial MATPOWER \cite{Matpower} $141$ bus distribution network used as the demonstration case. The distribution network is separated into $7$ aggregator zones ($N = 7$). The choice of $N$ and distribution of buses within $N$ is selected to increase common state variables between aggregators and to increase the number of 3-cliques in the communication graph ($\mathcal{G}_c$ in \Cref{fig:case141}) to satisfy the required conditions introduced in \Cref{sec:placement_measurement}. The utility is represented as the substation and belongs to the first aggregator. A total of 3900 prosumers are placed randomly across the distribution network. The total number of prosumers is arbitrary selected and sufficiently high to show algorithm scalability. The distributed pricing algorithm is run for six(6) ten-minute intervals to create a one-hour look-ahead window as a common duration of interest. The choice of interval length and the number of intervals can be selected to match local or regional guidelines on settlement time frames since the generalized formulations are independent of the length and number of intervals.

\subsection{Distributed Pricing}
\begin{figure}
    \centering
    \includegraphics[scale=1.3,keepaspectratio]{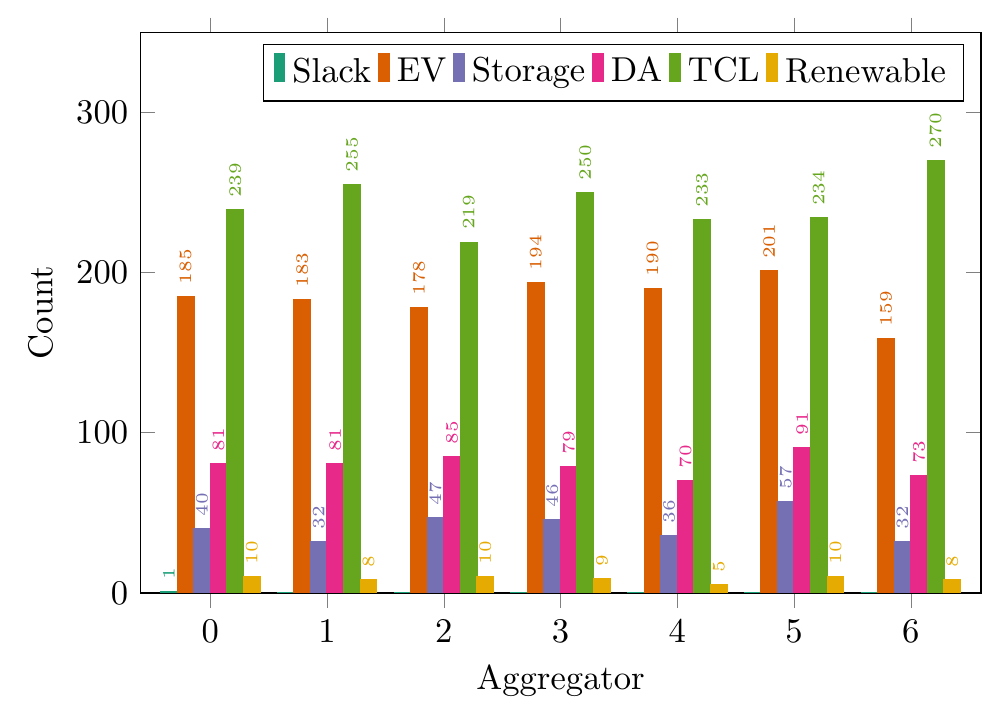}
    \caption{Asset types within each aggregator region.}
    \label{fig:assets}
\end{figure}
Each aggregator includes EVs, energy storage devices, DAs, thermostatically controlled loads (TCLs), and renewables, as summarized in \Cref{fig:assets}, with each aggregator including a similar percentage of each prosumer type (generated randomly). Each prosumer has different properties and cost/utility functions uniformly sampled from an identical distribution. Renewables are configured as price-takers (they have no cost/utility), TCLs have a quadratic cost function demanding payments for deviations from their thermostat reference temperature, storage devices require a linear payment proportional to their usage, and DAs and EVs have a linear cost function. The slack bus has a quadratic cost function. 

Cost/utility functions and prices are represented by an arbitrary monetary unit in which the price reflects the marginal cost of increasing load by a \textquote{single unit}. Any changes in asset cost parameters are expected to influence the converged price of \cref{eq.bus-problem}.
\begin{figure}
	\centering
	\subfloat[]{%
	\label{fig:power-resource}%
	\includegraphics[scale=1]{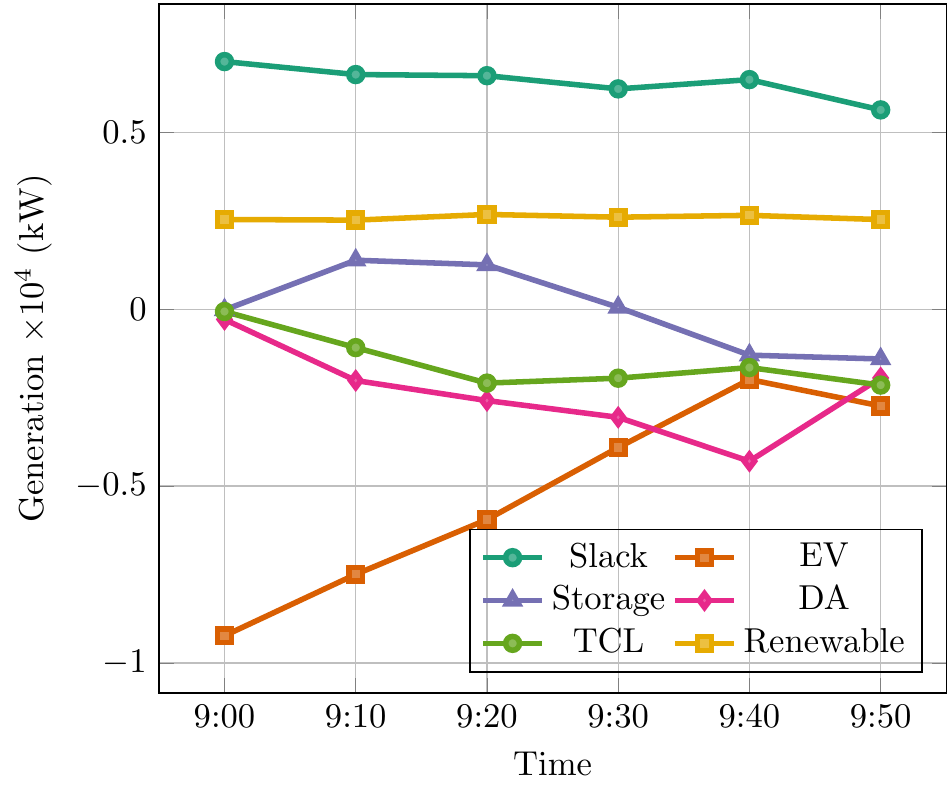}
	}%
	\\
	\subfloat[]{%
	\label{fig:price}%
	\includegraphics[scale=1]{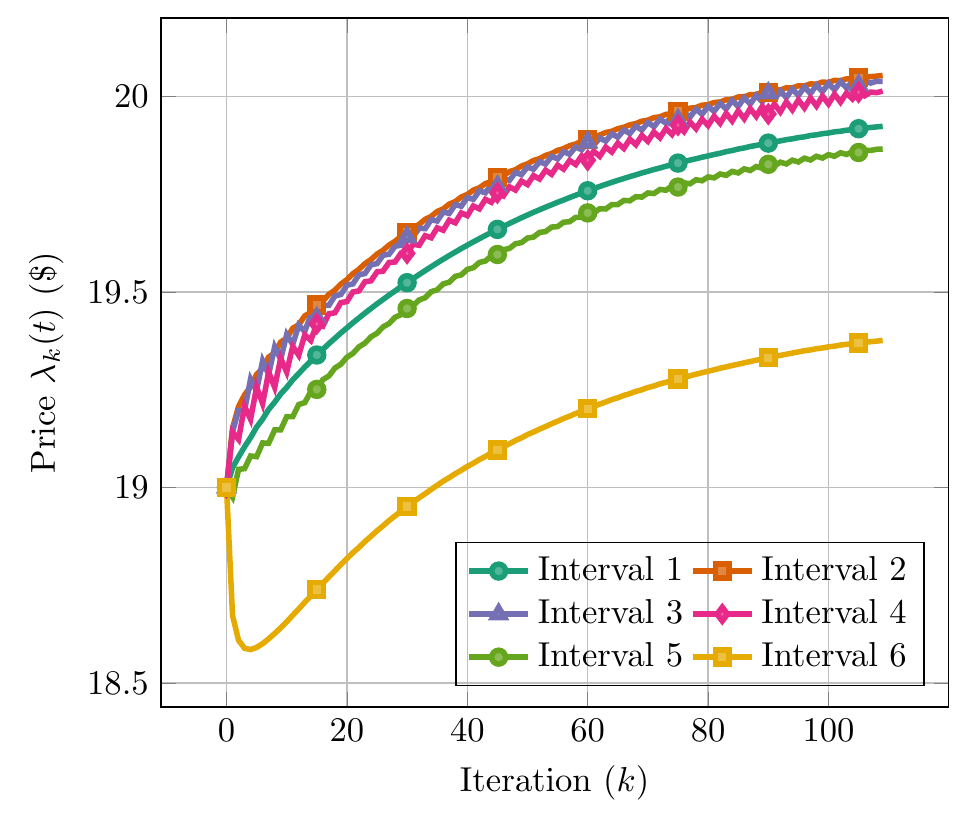}%
	}%
	\caption{\protect\subref{fig:power-resource} Power transfer between different resources \protect\subref{fig:price} Progression of the shadow price $\bm{\lambda}_k$ as Algorithm \ref{alg:distr-pricing} iterates, with each line denoting an element of $\bm{\lambda}$ which reflects price for a particular interval.}
	\label{fig:numericals1}
\end{figure}
Power transfer between different resources is shown in Figure~\ref{fig:power-resource}. The slack bus is a net generator (as the utility) since all other aggregators are composed mostly of consuming loads. Figure \ref{fig:price} shows prices for each time interval in the one-hour look-ahead window and how those prices evolve as the distributed algorithm iterates to solve all dispatches and prices for all intervals at once. The price only fluctuates by approximately 2\% across the iterations because of the significant load shifting behavior of flexible resources. A total of 28 out of 3900 prosumers are budget-constrained (need to curtail consumption due to insufficient funds); however, the algorithm converges without problems, even though convergence is not guaranteed.

To summarize, given a reasonable initial price for the algorithm, typically obtained from the previous solution, the algorithm solves in a distributed fashion the optimal schedule of all aggregators in a small number of iterations ($<100$ in this illustrative example) while managing to honor the individual budget constraint. As mentioned in Section \ref{sec.distributed}, convergence is not guaranteed due to the non-convexity introduced by the budget constraint in \eqref{eq:calpb}, which may lead to cyclic behavior in the gradient descent. However, for a low number of active constraints, the method often works as was the case in our experiments. This approach chooses an epsilon (Step \ref{step:ep_alg_2} in \Cref{alg:distr-pricing}) that expresses the price difference between iterations to determine when prices have stabilized, and hence iterations conclude. 

The approach developed here for distributed pricing can easily be adapted to the scenario when each prosumer can trade power directly rather than working through an aggregator. In that case, each prosumer will participate in the iterative pricing algorithm by interacting with the blockchain architecture through smart contracts as shown in Section 6.2. And while that approach is possible with the generalized mathematical framework introduced here, the authors advise caution as the approach will be computationally burdensome because the number of market participants engaging with the $GL$ has an exponential effect on the number of transactions and hence slows the convergence process.

\subsection{Distributed Verification}
Verification algorithm \Cref{alg:state_verification,alg:Security} presented in \Cref{sec:verification} are demonstrated here with the following parameters: $\epsilon_\pi = 1e-3$, $\epsilon=1e-3$, $\alpha_k = 1/k$, $c_3=c_4=0.5$, $\beta_n=2,~\forall n \in \mathcal{N}$. Measurements of available variables, $\bm{x}_{\mathcal{A}}$, were noisy versions of MATPOWER power flow output for the $141$ bus radial distribution feeder case. The noise, $w_n$, was chosen to be Gaussian with zero mean and a variance of $1$.

\begin{figure}
	\centering
	\subfloat[]{%
	\label{fig:noattack}%
	\includegraphics[scale=1.2,keepaspectratio]{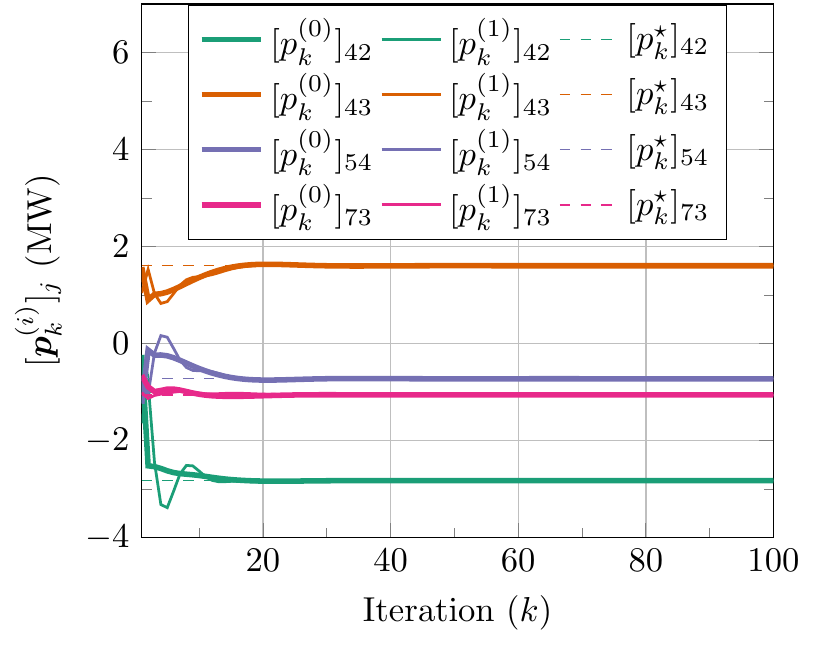}%
	}%
	\\
	\subfloat[]{%
	\label{fig:attack}%
	\includegraphics[scale=1.2,keepaspectratio]{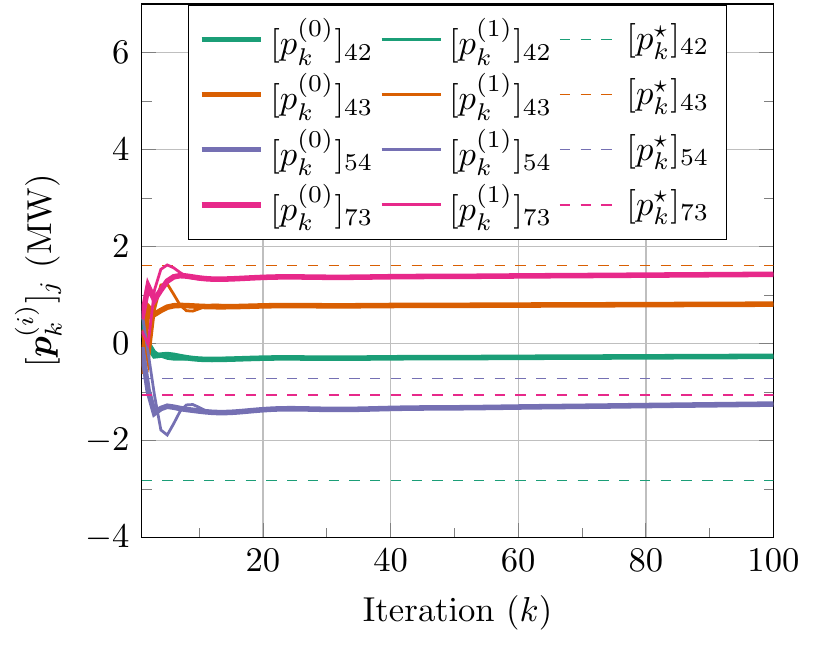}%
	}%
	\caption{Real power injections (in MW) by common nodes of aggregators $0$ and $1$ \protect\subref{fig:noattack} Convergence to the optimal point under no attack \protect\subref{fig:attack} Convergence to a non-optimal point when aggregator $0$ is an attacker.}
	\label{fig:numericals2}
\end{figure}
\begin{figure}
    \centering
    \includegraphics[scale=1.1]{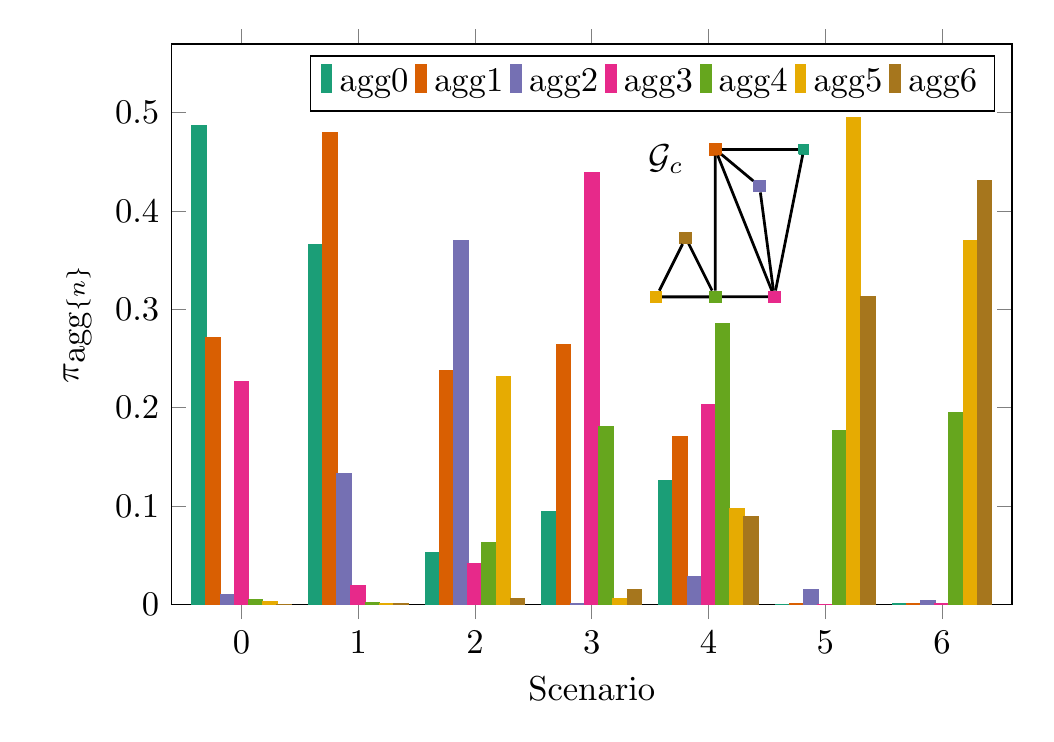}
    \caption{The stationary distribution, $\bm{\pi}$, of $\bm{B}$ under FDI attack. $\bm{\pi}$ represents the trust score: higher the value of $\pi_{\text{agg}\{n\}}$, lower is the trust in aggregator $n$. In scenario $m$, agg$\{m\}$ is the attacker.}
    \label{fig:stat_dist}
\end{figure}
To illustrate the veracity of the verification algorithm, we set one of the aggregators, $m$, to be a malicious entity capable of injecting false data into its communications with neighboring aggregators. The FDI attack from aggregator $m$ constitutes an injection from attack vector $\bm{a}^{(mn)}$ to the communications received by a neighboring aggregator $n$ of the attacker in \cref{eq:ADMM_phi}. In each iteration of the algorithm, the attack vector is chosen randomly subject to $\norm{\bm{a}^{(mn)}}_2 = 0.5\sqrt{\abs{\bm{S}_{mn}\bm{x}^{(m)}(t)}}$, where $\abs{\bm{S}_{mn}\bm{x}^{(m)}(t)}$ is the length of the vector $\bm{S}_{mn}\bm{x}^{(m)}(t)$, i.e., the number of common variables shared by region $m$ and its neighbor $n$. 

Consider an example with aggregators $0$ and $1$ that has buses 42, 43, 54, and 73 as adjacent nodes with common variables between two aggregator regions as shown in the network topology of \Cref{fig:case141}. Common variables include real power injection, reactive power injection, and voltage magnitude. \Cref{fig:noattack} shows convergence of the four sets of common variables (one for each bus) shared by aggregators when there is no attack, whereas \Cref{fig:attack} shows their divergence when region $0$ is an attacker. In \Cref{fig:noattack}, both aggregators converge to the optimal point $p^\star_b,~\forall b \in \{42,43,54,73\}$. The attack in \Cref{fig:attack} creates a situation in which the parameters reach to the same value yet do not converge at the optimal point. This prevents the verification algorithm from completing. The subroutine described in \Cref{alg:Security} aims to stop such attacks from occurring by tracking disagreements in the common variables between neighbors and identifying the most likely attacker.

The detection of FDI attacks from selfish entities on the TE market is accomplished using the stationary distribution $\bm{\pi}$ of the disagreement matrix $\bm{B}$, as discussed in \Cref{sec:attac_detect}. \Cref{fig:stat_dist} shows results for $N = 7$ in which each aggregator is shown to be the attacker. Each set contains seven bars representing the element of vector $\bm{\pi}$ corresponding to each of the seven aggregators. As the height of the bar increases, that aggregator is seen as more untrustworthy by the other aggregators. For example, in scenario $0$ where agg$0$ is the attacker, the corresponding bar plot indicates that the network of aggregators trust agg$0$ the least (i.e., $\pi_{\text{agg}0}$ is highest, as calculated in~\cref{algstep:pi_calc} of~\Cref{alg:Security}). Similarly, in the other scenarios, we observe that the corresponding attacker amasses the lowest trust level. 

It is worthwhile to discuss the distribution of distrust in the network. The distrust is spread among aggregators 0, 1, and 3 considering scenario $0$. This can be explained by reflecting on node connectivity of the $\mathcal{G}_c$ graph in \Cref{fig:stat_dist}, in which agg$1$ and agg$3$ are neighbors of agg$0$. In scenario $4$, where agg$4$ is the dishonest entity, we note that agg$4$ is the most untrustworthy, yet it is not as untrustworthy as the attackers in other scenarios. We also notice that distrust is more evenly spread across all aggregators in this scenario. In referring to $\mathcal{G}_c$, the reason for this outcome could occur because aggregator $4$ has the highest betweenness centrality and is the only \textit{cut vertex} in the network\footnote{A vertex in an undirected connected graph is a cut vertex iff removing it (and edges through it) disconnects the graph or creates more components than the original graph.}, i.e., agg$4$ controls information flow between the two clusters (aggregators 0, 1, 2, 3; and aggregators 5, 6). For the communication graph $\mathcal{G}_c$, agg$4$ dictates the spread of disagreements amongst the aggregators from either cluster. This process to identify attackers will then permit restart of the algorithm to complete the verification process. As billing then occurs the guilty party is penalized, fined, or disconnected from participating in the transactive energy network. 

\section{Design and Implementation on HLF}
\label{sec:imp}
This section details implementation on HLF of the CPS described in \Cref{sec:cps}, using the pricing algorithm from \Cref{sec:pricingAlgo}, and the verification algorithm in \Cref{sec:verification}. The choice of a permissioned blockchain architecture, such as HLF, allows consensus protocols that are far less energy-intensive than the proof-of-work consensus protocols employed by permissionless blockchain architectures \cite{wang2018blockchain}.

\subsection{Network Setup}
The ordering service for any blockchain framework requires a consensus protocol to ensure unambiguous ordering of transactions and guaranteed integrity and consistency of the blockchain across distributed nodes. The developed framework is adaptive to the currently supported Solo, Apache KAFKA, and RAFT algorithms for withstanding crash faults as an ordering service. 

Each aggregator is assigned a unique Certificate Authority (CA) \cite{hlf_paper} that is responsible for dynamically generating certifications (identities) for authenticating prosumers under each aggregator's purview. When joining the network, an individual prosumer shares information on their type of distributed energy resource and requests a new set of credentials with a unique prosumer identification (ID) from their associated aggregator. The credential issued by the aggregator includes a resource specific attribute in addition to the unique ID. Unlike prior works \cite{Saifur2018}, here the ID is embedded in the prosumer's certificate to add additional security to the issued certificate for that prosumer. Any interaction between the blockchain network and an aggregator requires admin identities \cite{hlf_paper}. In contrast, any communication between EVE and a prosumer involves the use of prosumer's unique identity generated by the respective CA.

The use of \textquote{channels} here provides the required isolation between individual aggregators and prosumers~\cite{channel_blog}. Each transaction in EVE is channel-specific and no data can pass among channels, ensuring privacy and efficient handling of parallel transactions. The EVE blockchain uses $N + 1$ number of separate channels (one channel for each of the $N$ individual aggregator for accessing the corresponding $LL$, plus one commonchannel among aggregators accessing the $GL$) to handle access to smart contracts and the ledger. A specific transaction with a specific ledger requires invoking the appropriate smart contract. 


\subsection{Implementation of EVE Through Smart Contracts}
\label{sec:smart_contracts}
\begin{figure*}[htbp]
	\centering
    \includegraphics[width=\textwidth]{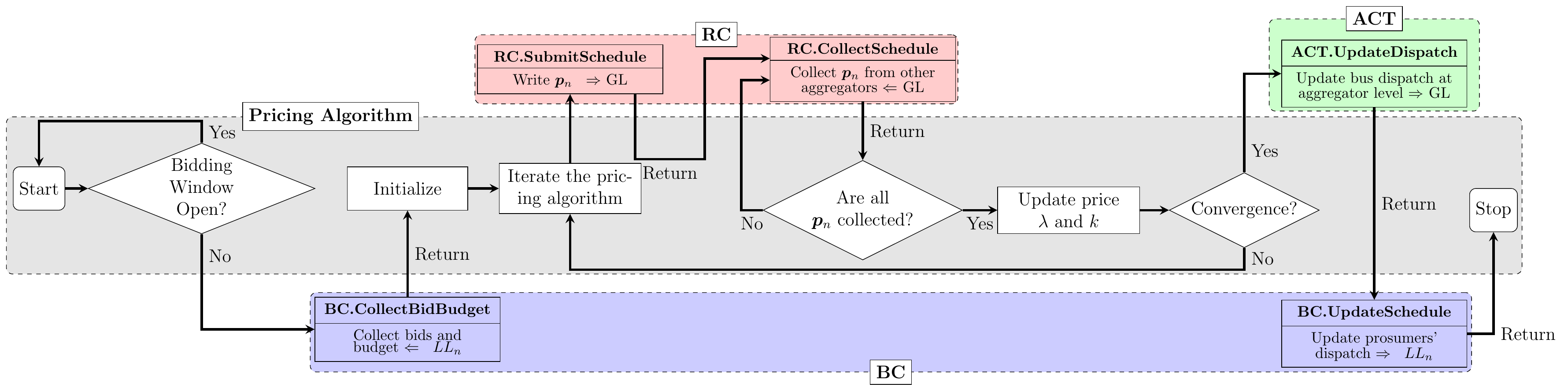}
	\caption{Graphical depiction of pricing Algorithms \ref{alg:pricing-prosumer} and \ref{alg:distr-pricing} using smart contracts and ledgers for aggregator $n \in \mathcal{N}$.}
	\label{fig:EVE_implementationa}
\end{figure*}

\begin{figure*}[htbp]
	\centering
    \includegraphics[width=\textwidth]{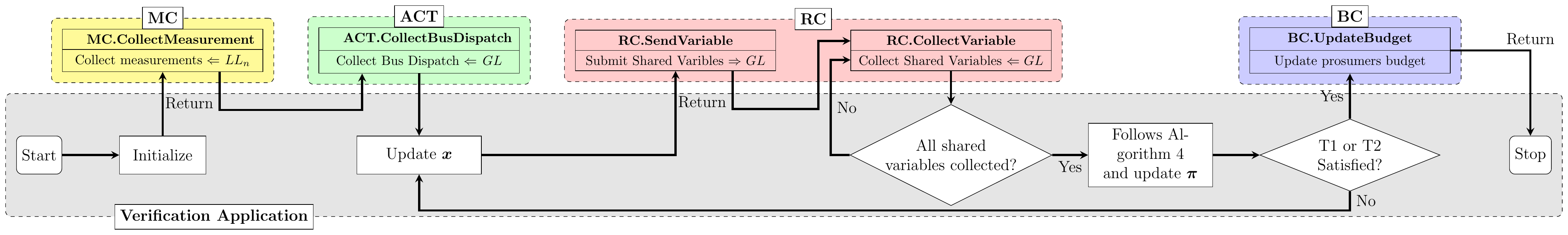}
	\caption{Graphical depiction of verification \Cref{alg:state_verification} using smart contracts and ledgers for aggregator $n \in \mathcal{N}$.}
	\label{fig:EVE_implementationb}
\end{figure*}

\Cref{tab:channel_description} lists associations between channels, smart contracts, ledgers, and participant access control. Interactions between smart contracts is summarized in \Cref{fig:EVE_implementationa} and \Cref{fig:EVE_implementationb} for pricing and verification, respectively. Pricing and verification algorithms are written in Python as external applications. \texttt{Node.js} applications are developed to handle communications between external applications and smart contracts allowing read and write operations to appropriate ledgers. Note that each application interacts with different smart contracts to accomplish its required objective.

\begin{table}
\caption{List of channels and associated smart contracts, ledgers, and participant access for $\mathcal{N} = \{0,\dots,6\}$.}
\label{tab:channel_description}
\resizebox{\columnwidth}{!}{%
\begin{tabular}{|c|c|c|c|}
\hline
Channel & \makecell{Installed Smart Contract} & \makecell{Ledger  Access} & \makecell{Participant Access} \\ \hline
commonchannel        &       \textbf{Account Contract, Record Contract}                   &      $GL$                        &   aggregator $n;~\forall n \in \mathcal{N}$                 \\ \hline
agg\{$n$\}channel          &       \textbf{Bid Contract, Measurement Contract}                   &      $LL_n$       &   \makecell{aggregator $n \in \mathcal{N}$ and \\ all prosumers $b\in\mathcal{B}^{(n)};~n \in \mathcal{N}$ }  \\ \hline
\end{tabular}%
}
\end{table}

\textit{\textbf{Account Contract (ACT)}} manages aggregator-level information such as total production/consumption and associated total cost at the bus-level. At the beginning of the bidding window, each aggregator is required to establish their respective bus information by creating timestamped entries in the $GL$ through \textbf{ACT}.

\textit{\textbf{Bid Contract (BC)}} contains a set of functions to manage individual prosumer bids, dispatch values, and budget information. This contract allows a prosumer to use its certificate to submit any number of bids within the bidding window. However, only the last submitted bid within the bidding window is accepted by each aggregator. When submitting a bid, the prosumer is required to provide its asset type (i.e., EV, renewable generation, DA, TCL, storage device, and inflexible load) and corresponding parameters. The smart contract extracts the prosumer's ID from the certificate and ties the submitted bid with the prosumer's unique identity to prevent malicious prosumers from impersonating other prosumers. After the bidding window closes, the aggregator queries submitted bids with their associated budgets from the $LL$ through the aggregator's dedicated channel. The aggregator then executes the distributed pricing algorithm by exchanging information with other aggregators.

\textit{\textbf{Record Contract (RC)}} is responsible for data exchange for the iterative pricing algorithm through $GL$. After achieving convergence, each aggregator updates its prosumers' dispatch and cost information in the $LL$, and bus dispatch values in the $GL$ for future verification. This iteration can be computationally intensive because each interaction (reading and writing) with the $GL$ is a transaction in HLF. 

\textit{\textbf{Measurement Contract (MC)}} handles local measurements from smart meters installed within individual aggregator zones for future verification purposes. Note that the developed algorithm is independent of the sensor location, allowing smart meters to be placed randomly in each aggregator zone for illustration purposes here. The EVE framework can accept measurements at each interval of $\mathcal{T}$ as separate transactions or all measurements for $T$ intervals at the same time as a single transaction. The distributed verification step is always one time step behind the distributed pricing algorithm. For the simulated framework, data from smart meters are generated by solving a non-linear AC power flow problem using the prosumers' dispatch values as input and then adding noise to it. The \textbf{RC} handles information exchange for the distributed verification process. Each transaction is assigned a \texttt{type} to ensure separation of entries inside \textbf{RC} for pricing and verification, thus allowing the use of a single smart contract to handle both iterative algorithms.  

\Cref{alg:distr-pricing,alg:state_verification} have different data sharing requirements. Each iteration of \Cref{alg:distr-pricing} by one aggregator requires information from all aggregators, whereas each iteration of \Cref{alg:state_verification} requires information from just neighboring aggregators. Private data sharing between neighboring aggregators can be achieved by creating neighbor-specific channels. Nevertheless, this process is burdensome because 1) more channels are required to handle private data sharing, and 2) changes are required in the existing blockchain network if the communication graph changes (due to changes in sensor deployments among aggregators or distribution network reconfiguration). Therefore, the developed framework leverages the ABAC feature of HLF to handle private communication using the existing commonchannel. When a new edge is created in the communication graph, the new relationship can be added to \textit{RC} by upgrading the smart contract following the rules of the initial agreement.

\begin{table}
\centering
\caption{Benchmark Results for Hyperledger Caliper to Test $200$ Iterations of Information Exchange for Algorithm \ref{alg:distr-pricing}.}
\label{table:recordcaliper}
\resizebox{0.98\textwidth}{!}{%
\begin{tabular}{|c|c|c|c|c|c|}
\hline
$N$ & \makecell{Total \\ Transactions} & \makecell{Sent \\ Rate (tps)} & \makecell{Max \\ Latency (s)} & \makecell{Min \\ Latency (s)} & \makecell{Throughput \\ (tps)} \\
\hline
6 & 1200 & 6 & 1.8 & 0.31 & 6 \\ \hline
7 & 1400 & 7 & 1.55 & 0.32 & 7 \\ \hline
8 & 1600 & 8 & 1.77 & 0.36 & 8 \\ \hline
9 & 1800 & 9 & 1.59 & 0.38 & 9 \\ \hline
10 & 2000 & 10 & 1.49 & 0.39 & 10 \\ \hline
\end{tabular}%
}
\end{table}
\subsection{Results}

\textcolor{black}{Performance results of the proposed framework are generated using predefined use cases in Hyperledger Caliper. For test cases with varying numbers of aggregators ($N$) between $6-10$, and assuming the pricing algorithm requires $200$ iterations for convergence, \Cref{table:recordcaliper} shows benchmark results in terms of the maximum latency,  minimum latency, and throughput\footnote{Latency = (time when response received - submit time) in second. Throughput = (total valid committed transactions / total time in seconds for all committed nodes in the network) in transactions per second (tps).}. For each scenario shown in \Cref{table:recordcaliper}, the sending tps rate is equivalent to $N$ because the number of aggregators is the maximum number of writes that can occur to the ledger.} 

\subsection{Security Analysis}

\begin{table}
\caption{Security Analysis of Reviewed Surveys.}
\label{tab:survey}
\resizebox{\columnwidth}{!}{%
\begin{tabular}{|c|c|c|c|c|c|c|}
\hline
Reference & EVE & \cite{Wang2019,Saifur2018} & \cite{gai2019privacy} & \cite{laszka2018transax} & \cite{Jonathan2018,peer_to_peer,munsing2017blockchains} & \cite{danzi2017distributed} \\ \hline
\makecell{Implementation \\ Framework} & \multicolumn{3}{c|}{Hyperledger} & \multicolumn{3}{c|}{Ethereum} \\ \hline
Threat Model & \checkmark & x & \checkmark & \checkmark & x & \checkmark \\ \hline
Attack Scenario & \checkmark & x & \checkmark & x & x & \checkmark \\ \hline
Physical Verification & \checkmark & x & x & x & x & x \\ \hline
\end{tabular}%
}
\end{table}

Security and privacy of the proposed EVE blockchain introduced here are compared against other leading blockchain approaches. \Cref{tab:survey} summaries security considerations of the reviewed works that used blockchain for TE. Most do not fully consider security aspects such as threat model, attack scenario, and verification mechanism at the physical layer. This indicates that the security of these works depends on the built-in security mechanism of the blockchain framework and is not analyzed in-depth for additional threats or weaknesses. In this work, the proposed EVE blockchain framework has been designed and implemented with direct inclusion of cyber-security and specific threat models and attack scenarios. \Cref{tbl:feasible} provides a summary of potential threats and countermeasures referenced by the NIST Guide to Industrial Control Systems~\cite{stouffer2015guide} and the HLF security mechanism as they relate to the threat model outlined in \Cref{sec:malbeh}.

As the currently supported ordering mechanisms in HLF only provide crash fault tolerance and do not provide BFT \cite{Fabric_consensus}, a customized BFT-SMART~\cite{sousa2018byzantine} state machine replication and a consensus library has been integrated into this work too. By reinforcing the design with this mechanism, our approach can achieve BFT resilience and durability to avoid a single point of failure. Also, using sensor measurements in the verification stage allows EVE to assure prosumers' compliance with the scheduled transaction.

\begin{table}
\centering
\caption{Feasible Threats and Countermeasures in EVE.} 
\label{tbl:feasible}
\resizebox{\textwidth}{!}{%
\begin{tabular}{|c|c|c|c|}
\hline
\textbf{Layer} & \textbf{Feasible Threats}  & \textbf{Countermeasures (HLF)}  & \textbf{Countermeasures (EVE)}  \\ \hline
Application & \makecell{Stealth FDIA, \\ DoS attack, \\ Smart Contract, \\ Malware} & MSP (Fabric CA) & \makecell{MSP, \\ FDIA Detection \\ (Physical Verification)} \\ \hline
Blockchain & \makecell{Relay attacks,\\ Privilege Elevation,\\ Repudiation,\\ Info disclosure,\\ Byzantine Fault,\\ Civil attack} & \makecell{Read/Write Set Validation, \\ MSP Tracibility with\\ digital signature,  \\ Channel isolation}   &  \makecell{BFT-SMART \\  with features \\ from HLF} \\ \hline
Network & \makecell{DoS attack, \\ Eclipse attack \cite{heilman2015eclipse}} & TLS & \makecell{ABAC with features \\ from HLF} \\ \hline
Client & \makecell{Identity Theft, \\ Malware} & \makecell{MSP (Fabric CA), \\ Hardware  Security \\ Module} &  \makecell{ABAC with features \\ from HLF}\\ \hline
\end{tabular}%
}
\end{table}

\section{Conclusion}
A blockchain-enabled transactive energy platform entitled \textbf{Electron Volt Exchange} is presented in this paper. The integration of blockchain allowed a secured process for handling individual bids (prosumers) and collective bids (aggregators). Implementing aggregators for the distributed pricing algorithm allowed efficient use of the Hyperledger Fabric distributed architecture, as demonstrated here for a $141$ bus radial network. A secure mechanism for pricing and later verifying economic transactions through a distributed consensus process is also presented. Future work will explore the implementation of additional Hyperledger Fabric features (e.g., idemix, smart contract packaging), other market mechanisms, and verification algorithms through distributed consensus for meshed networks.

\section*{Acknowledgement}
This study was funded in part by the United States Office of Naval Research (ONR) Defense University Research-to-Adoption (DURA) Initiative under award number N00014-18-1-2393.


\appendix
\section{Demand Response Resource Models}\label{app:drmodels}
\subsection{Electric Vehicles (EVs)}
An EV requires a certain amount of charge $\overline{u}$ over a period of length $\tau$. Primary constraints are: 
\begin{enumerate}[leftmargin=*]
  \item Rate of charge when grid-connected is $\dot u(t)=-p(t)$ for $0\leq u(t)\leq \overline{u}$ and zero when the EV is full, with $p(t)<0$ since charging is a load.
  \item Charging is permitted only at a constant rate of $-\rho$, i.e.,\ $p(t)\in\{0,-\rho\}$. Discharge to the grid is not permitted. 
  \item Charging has a deadline, i.e.,\ $u(t_d)=u(t_a+\tau_c+\tau_s)=\overline{u}$. 
\end{enumerate}
Here $t_a$ denotes the EV arrival time, $t_d$ the EV departure time, $\tau_c$ the time needed for charging, and $\tau_s$ the leftover (slack) time.
Considering a large number of loads, the constraint on charging rate can be relaxed as
$ -\rho \leq p(t)\leq  0 $. 
In discrete time and vector form, assuming that the variable $u(t)$ is the energy normalized by the sampling period (i.e.,\ the time that elapses between $t$ and $t+1$), we can write:
\begin{equation}\label{eq.ev}
  \bm{p}=\bm{A}\bm{u}+\bm{\ell},~ u(t_d) = \overline{u},  ~-\rho \leq p(t)\leq  0~\forall~t \in \mathcal{T}
\end{equation}
In \cref{eq.ev}, $\bm{A}$ computes the finite difference of the state of charge values in $\bm{u}$. Using $\bm{J}$ to denote an off-diagonal shift matrix, the following can be written:
\begin{equation}
  \bm{A}=(\bm{J}-\mathbb{I})\in \mathbb{R}^{|\mathcal{T}|\times|\mathcal{T}|},~\bm{\ell}=[u(t_a), 0, \ldots, 0]^\T \in \mathbb{R}^{|\mathcal{T}|\times 1}
\end{equation}
In this case $\bm{A}^{\dagger}=\bm{A}^{-1}$ is a triangular matrix of all negative ones. $\bm{A}^{\dagger}$ performs a cumulative sum of the entries of $-\bm{p}$, which is the inverse operation of taking the finite difference (similar to the integral is the inverse operation of the derivative). In general, we assume a prosumer is willing to pay more for having their EV charged earlier, and expects a discounted electricity price if not charging at full power. This conceptualization of energy price is a function of time, with price decreasing monotonically with time as $\bm{c}_1$ and $c_\text{d} < [\bm{c}_1]_t~\forall t\in\mathcal{T}$:
\begin{subequations}
\begin{align}
  \!\!\! C(\bm{p}) &= \bm{c}_1^\T \bm{p} - c_\text{d}\textnormal{min}(\overline{u} - u_{|\mathcal{T}|}, \rho\, \textnormal{max}(0, \tau_c+\tau_s-|\mathcal{T}|))\\ 
      &= \bm{c}_1^\T \bm{p} - c_\text{d}\textnormal{min}(\overline{u} + \bm{1}^\T\bm{p}, \rho \, \textnormal{max}(0, \tau_c+\tau_s-|\mathcal{T}|))
\end{align}
\end{subequations}

\subsection{Deferrable Appliances (DAs)}
These loads include appliances such as washers, dryers, and water pumps that can be programmed to start their cycle at different times of day. The feasible set of power demand is based on a load profile $h(t)$, a minimum activation time $0\leq t_a\leq \abs{\mathcal{T}}-1$, and a slack $\tau_s \geq 0 $: 
\begin{equation}
  p(t)=-h(t-t_a-\tau), ~~~0\leq \tau\leq \tau_s
\end{equation}
where the price function depends on the slack $\tau_s$. Let us assume without loss of generality that $t_a=0$ as the arrival time can be embedded into the the signal $h(t)$. 
Suppose also that the duration of $h(t)$ is $d$ and that the slack and time are discrete. The signal $u(t)$ can indicate the time at which DA starts its cycle, which naturally means that $u(t)\in\{0,1\}$ and the $\ell_1$ norm $\|\bm{u}\|_{1}=1$. 
Considering a large enough population, this constraint can be relaxed to obtain an approximation of the feasible set as follows:
\begin{equation}
  \bm{p}=\bm{A}\bm{u},~~\|\bm{u}\|_1=1
\end{equation}
where $\bm{A} \in \mathbb{R}^{\left(|\mathcal{T}|+d\right) \times |\mathcal{T}|}$ equal to:
\begin{equation}
    \bm{A}^\T=-
    \begin{bmatrix}
    h(0)& \dots & h(d)  &   0  &\dots &  0\\
    0   &  h(0) &\ldots & h(d) &\dots & \vdots\\
    \vdots & \ddots &  \ddots &  \ddots & \ddots &\vdots\\
     0 & \dots &  0 &  h(0) & \dots &h(d)
    \end{bmatrix}
\end{equation}
The price a prosumer is willing to pay decreases as the delay increases. The price ranges from a maximum price the consumer is willing to pay to a minimum price that is the lowest possible energy cost. Suppose that we want the cost to grow linearly with time, and let $c$ be the constant in the cost expression. Let $\tilde{\bm{c}}_1=\tilde{c}_1\cdot(|\mathcal{T}|-1,|\mathcal{T}|-2,\ldots,1,0) $. The cost can be obtained as follows:
\begin{align}
    C(\bm{p})=\bm{c}_1^\T\bm{p}+c_0= \tilde{c}_1\sum_{t=0}^{|\mathcal{T}|-1}(|\mathcal{T}|-1-t) \cdot u(t)+\tilde{c}_0,\\
    \Rightarrow~~\bm{c}^\T_1=\tilde{\bm{c}}_1\bm{A}^{\dagger},
    ~~c_0=\tilde{c}_0
\end{align}

\subsection{Thermostatically Controlled Loads (TCLs)}
These loads include space heaters, air conditioners, and water heaters. Similar to prior works \cite{callaway2009tapping}, we assume that the temperature dynamics of a heat pump based TCL can be modeled as a first-order differential equation: 
\begin{equation}
  C \dot \theta(t)= (\theta_\text{o}(t) - \theta(t))R^{-1}+p(t)\eta  +\varepsilon(t)R^{-1},~~~ 
  p(t) \in \{0,-\rho\}\label{eq:de}
\end{equation}
with $R$ being thermal resistance, $C$ thermal capacitance, $\theta(t)$ the inside temperature, $\theta_\text{o}(t)$ the outdoor temperature, $\eta$ the efficiency of the heat pump ($\eta>0$ for cooling and $\eta<0$ for heating), $\rho$ continuous electrical power rating \footnote{Water heaters can be described using the same principles, with an additional energy loss component describing the hot water being replaced by cold water. However, in this paper, we will focus on heat pump based TCL because they are more dependent on external temperatures than water boilers.}, and $\varepsilon(t)$ denoting a random perturbation of temperature by external factors such as opening of windows/doors or operation of stoves. We denote the thermostats reference temperature as $\theta_\text{r}$. 
Let:
\begin{equation}\label{eq.defTCL}
\begin{aligned}
u(t)&\triangleq \frac{\theta(t)-\theta_r(t)}{ R\eta}\\
\tilde{\ell}(t)&\triangleq \frac{\theta_r(t)-\theta_o(t)}{R\eta}+\frac{\varepsilon(t)}{R\eta}+\frac{C}{\eta}\dot{\theta}_r(t),~~
\tau_h\triangleq CR
\end{aligned}
\end{equation}
To express the inter-temporal constraints as well as the comfort zone of the user, we can rearrange and relax the set of constraints in \cref{eq:de} as follows:
\begin{subequations}
\begin{align}
    \label{eq.TCLp}
    p(t) &= \frac{C}{\eta} (\dot{\theta}(t)-\dot{\theta}_r(t))+\frac{\theta(t)-\theta_r(t)}{R\eta}+
    \frac{\theta_r(t)-\theta_o(t)}{R\eta}+\frac{\varepsilon(t)}{R\eta}+\frac{C}{\eta}\dot{\theta}_r(t)\\
    &=\tau_h \dot{u}(t)+u(t)+\tilde{\ell}(t),~\underline{u}\leq u(t)\leq \overline{u}, ~
 -\rho\leq p(t)\leq 0 
\end{align}
\end{subequations}
where we have relaxed the integer constraint $p(t) \in \{0,-\rho\}$.

It is notable that the $\tilde{\ell}(t)$ term in this affine relationship is random, since the outdoor temperature is random and so is the perturbation of the indoor temperature from sources other than the heat pump. 
In discrete time, we can average the behavior and approximate the relationship as follows:
\begin{equation}
  \bm{p}= \bm{A}\bm{u} + \bm{\ell},~~~\underline{u}\leq u(t)\leq \overline{u},~~~  -\rho\leq p(t)\leq 0 
\end{equation}
Using \cref{eq.defTCL,eq.TCLp} the following can be written:
\begin{equation}
\begin{aligned}
&\bm{A}=\tau_h(\mathbb{I}-\bm{J})+\mathbb{I},~\bm{A}^{\dagger}=\bm{A}^{-1}=((\tau_h+1)\mathbb{I}-\tau_h\bm{J})^{-1}\\
&\bm{\ell}=\tilde{\bm{\ell}}-\tau_h [u(0),\bm{0}^\T]^\T \in \mathbb{R}^{|\mathcal{T}|\times 1}
\end{aligned}
\end{equation}
and $\tilde{\bm{\ell}}=(\tilde{\ell}(1),\ldots,\tilde{\ell}(|\mathcal{T}|))$ with $\tilde{\ell}(t)$ being defined in \cref{eq.defTCL}.
The price demand function should represent the prosumer's willingness to deviate from reference temperature. The prosumer is willing to pay less for larger temperature deviations than expected. We represent this cost as quadratic with the value of $\bm{u}$, i.e., proportional to $\norm{\bm{u}}^2$.
This means that the demand function is $\tilde{c}_0-\tilde c_2\|\bm{u}\|^2$ and:
\begin{equation}
    C(\bm{p})=\bm{p}^\T\bm{C}_2\bm{p}+\bm{p}^\T\bm{c}_1+c_0=\tilde{c}_0-\tilde c_2\norm{\bm{A}^\dagger(\bm{p}-\bm{\ell})}^2
\end{equation}
where:
\begin{equation}
   \rightarrow~~\bm{C}_2=-
   \tilde{c}_2\bm{A}^{-2},~\bm{c}_1=2\tilde{c}_2\bm{\ell}\bm{A}^{-2},~~c_0=\tilde{c}_0-
   \tilde{c}_2\bm{\ell}^\T\bm{A}^{-2}\bm{\ell}
\end{equation}
  
\subsection{Storage Devices}
Typically, battery rate of charge or discharge is constrained to be a constant value, meaning $p(t)\in\{-\rho,0,\rho\}$. If we relax this non-convex constraint as done before, the load constraints are analogous to that of an EV, except that the unit can discharge:
\begin{equation}\label{eq.battery}
  \bm{p}=(\mathbb{I}-\bm{J})\bm{u}-[u(0),\bm{0}],~~~\abs{p(t)}\leq \rho,~~0\leq u(t)\leq \overline{u}  
\end{equation}
where $\bm{J}$ is the shift matrix, shifting to the right each of the entries of $\bm{u}$.
Here it is natural to assume that the price for discharging is higher than the price of charging, but it is also possible to express a cost that depends on the battery state. In addition, if storage is charged by the random injection of, for instance, solar PV, the forecast can be incorporated into the vector $\bm{\ell}$ in the model.  We ignore other complexities here for simplicity. Considering $(a)_+=\max(0,a)$, we can express the cost as:
\begin{equation}
  C(\bm{p})=(\bm{p})_+^\T(\bm{c}_1^+)+(-\bm{p})_+^\T(\bm{c}_1^-)    
\end{equation}
where $\bm{c}_1^+$ and $\bm{c}_1^-$ are non-negative cost vectors.

\subsection{Renewables} 
The power injection from wind or solar PV has no marginal cost, therefore the only meaningful way for renewables to participate is posting a forecast of future production $\bm{p}$ with zero cost, i.e., $C(\bm{p})=0$. 

\subsection{Supply from the Transmission Grid} 
We assume that the transmission grid appears in the system as the slack bus and has a certain cost function for selling and a certain cost function for buying power below and above a schedule $\bm{p}_s$ that was cleared in previous wholesale market stages. 
Therefore, at the substation bus we have a single supplier with supply function:
\begin{equation}\label{grid}
  C(\bm{p})=(\bm{p}-\bm{p}_s)_+^\T(\bm{c}_1^+)+(-\bm{p}+\bm{p}_s)_+^\T(\bm{c}_1^-)       
\end{equation}
For simplicity we assume that any deviation is feasible, so that the dispatch is always feasible, and the slack bus compensates for any shortfall or surplus of power subject to physical constraints in \ref{app:StateVer}.

\section{Electric Grid Constraints}
\label{app:StateVer}
A radial electrical distribution system can be represented by a set of buses $\mathcal{B}$, edges $\mathcal{E}_e$, and a root node (commonly the substation node or slack bus) where each edge $\ell =(b,b')\in \mathcal{E}_e$ and $b,b'\in \mathcal{B}$. The \emph{from} and \emph{to} functions are defined as $f(\ell)=b$ and $t(\ell)=b'$ to return the source node and incident node for an edge, respectively.
The inverse function $t^{-1}(b)$ gives back the edge pointing to bus $b$ (for a radial graph the \textit{to} bus $b$ for each edge is unique) and $f^{-1}(b)$ returns all the edges originating at bus $b$. 
Dropping the time index for brevity, the power flow equations at each of the branches $\ell \in \mathcal{E}_e$ are given by \cite{barenwunetworkreconfig}:
\begin{subequations}
\begin{align}
    p_{t(\ell)}&=P_{\ell}-\sum_{\ell'\in f^{-1}(t(\ell))}P_{\ell'}-\Re(z_{\ell})c^2_{\ell}
    \label{eq:PF1}\\
    q_{t(\ell)}&=Q_{\ell}-\sum_{\ell'\in f^{-1}(t(\ell))} Q_{\ell'}-\Im(z_{\ell})c^2_{\ell}\label{eq:PF2}\\
    v_{f(\ell)}^2&=v_{t(\ell)}^2+
    2(\Re(z_{\ell})P_{\ell}+\Im(z_{\ell})Q_{\ell})-|z_{\ell}|^2c_{\ell}^2\label{eq:PF3}\\
    v^2_{f(\ell)}&=\frac{P^2_{\ell}+Q^2_{\ell}}{c^{2}_{\ell}}.\label{eq:PF4}
\end{align}
\end{subequations}
All these equations are linear in the bus and branch quantities $(p_b(t), q_b(t),\allowbreak v_b^2(t))$ and $(P_{\ell}(t),Q_{\ell}(t),c^2_{\ell}(t)
)$ except \cref{eq:PF4}. However, including an auxiliary variable 
$x'_{\ell}=\frac{P^2_{\ell}+Q^2_{\ell}}{c^{2}_{\ell}}$ can simplify the description of the physical constraints. By expressing $\bm{x}(t)$ as the auxiliary variables, then without loss of generality \crefrange{eq:PF1}{eq:PF4} can be written in the following linear form:
\begin{equation}\label{eq:linear-constr}
    \bm{H}\bm{x}=\bm{H}_{\mathcal{A}}\bm{x}_{\mathcal{A}}+\bm{H}_u\bm{x}_u=\bm{0}, 
\end{equation}
Though constraints in \cref{eq:elec_grid_const} are non-linear in general, here the constraints are relaxed by ignoring the non-linear relationships among the auxiliary variables and the remaining entries of the vector $\bm{x}$. The measurements are represented by  $\bm{z}=\bm{x}_{\mathcal{A}}+\bm{\epsilon}$ and the physics of the system implies that $\bm{x}_{\mathcal{A}}$ satisfies \cref{eq:linear-constr}. 

From the vantage point of each aggregator region $n$, only a subset of the variables $\bm{x}_{\mathcal{A}}^{(n)}$ are measured, meaning that $\bm{z}^{(n)}=\bm{x}_{\mathcal{A}}^{(n)}+\bm{\epsilon}^{(n)}$. 
Also, not all constraints in \cref{eq:linear-constr} include the variables $\bm{x}^{(n)}$. Hence, the equations that involve buses/lines in region $i$ can be isolated and written as:
\begin{equation}\label{eq:linear-constr_region}
\bm{h}^{(n)}(\bm{x}^{(n)})=\bm{H}^{(n)}\bm{x}^{(n)}=\bm{0}.   
\end{equation}
These equations are used to verify that measurements and injections values are consistent with the laws of physics. The available measurements in each zone and the neighboring zones are used to interpolate the unknown variables as discussed in \Cref{sec:RSVProblem}.

\bibliographystyle{elsarticle-num-names}
\bibliography{bibilography}

\end{document}